\pdfoutput=1
\documentclass[11pt,onecolumn]{article}

\usepackage[margin=0.9in]{geometry}
\usepackage{amsmath,amssymb}
\usepackage{graphicx}
\usepackage{booktabs}
\usepackage{array}
\usepackage{caption}

\usepackage[hidelinks]{hyperref}
\usepackage{orcidlink}
\usepackage{microtype}
\newcommand{\td}{d_{\mathrm{tr}}}
\newcommand{\bt}{\tilde{\beta}}
\newcommand{\betaeff}{\beta_{\mathrm{eff}}}
\newcommand{\twoq}{\ensuremath{n_{2\mathrm{q}}}}

\title{\textbf{A common-gate-set, accuracy-matched comparison\\
of quantum thermal-state preparation algorithms}}

\author{Tushar Pandey\,\orcidlink{0000-0001-7448-5723}}

\date{November 2026}

\begin{document}
\maketitle

\begin{abstract}
\noindent
Quantum thermal-state preparation algorithms report costs in incommensurable currencies: oracle queries,
Hamiltonian evolution time, or device gate depth. We show that the gap between an algorithm's own currency
and a realizable circuit is method-dependent: under second-order product-formula compilation it reaches
$3\times10^6$ for a double-bracket protocol on the critical transverse-field Ising chain and $2\times10^7$
on Heisenberg, against at most $1.7$ for analog-native local driving. We compile three
reimplemented protocols from one schedule with one compiler, so the counted circuit is the simulated
unitary, and match 36 cells in trace distance under five pre-declared fairness rules, two compilation
regimes and six cost models. The arms have three distinct accuracy floors, and the Heisenberg chain adds a
fourth: a jump family non-ergodic under SU(2) symmetry. On this grid, rankings invert with accuracy
target, gate budget and cost model, while the compilation regime reorders only near-ties. Reimplementation surfaces a
discrepancy in each source paper. Published-resource rankings are therefore biased in a known direction.
\end{abstract}

\section{Introduction}
\label{sec:intro}

Preparing the Gibbs state $\rho_\beta \propto e^{-\beta H}$ of a Hamiltonian $H$ at inverse temperature
$\beta$ is a central primitive for quantum simulation. The last three years have produced a large body of
algorithms, and three families dominate the near-term literature, each reporting its cost in a different
currency. We compare the three at a common gate set and matched accuracy, and we show that the gap between
each algorithm's own currency and a realizable circuit depends on the method: under second-order
product-formula compilation it reaches $3\times10^6$ for a double-bracket protocol on the critical
transverse-field Ising chain, against at most $1.7$ for local driving.

The first family is repeated interaction with a single ancilla.
Ding, Zhan, Preskill and Lin~\cite{ding2025simple} couple the system through a filtered, time-dependent
interaction to one ancilla that is reset every cycle. The fixed point of the resulting channel is the
Gibbs state, proved in the weak-coupling regime $\alpha \ll 1$, $\alpha T \ll 1$, with $\alpha$ the
coupling strength and $T$ the length of the interaction window. The single-ancilla
design builds on the Lindbladian ground-state construction of Ref.~\cite{ding2024single}. Its companion
experiment~\cite{farrell2026preparing} runs the protocol on 139 superconducting qubits and states every
hyperparameter: $\alpha = 1.75$, a bath-frequency cutoff $\omega_{\max} = 8$, $T = 0.75$, a Trotter step
$\delta t = 0.25$, three Trotter steps per cycle, a constant coupling envelope (filter width
$\sigma = \infty$), and $N_{\mathrm{resets}} = 7$ for its flagship run with $N_S = 79$ system and
$N_E = 60$ environment qubits. Its resource statement is a two-qubit-gate depth of 55 per cycle for a
kagome-lattice model, which
that experiment maps onto heavy-hex hardware with SWAP insertions rather than running on kagome
connectivity.

The second family is the double-bracket thermofield double (DB-TFD).
Wright \emph{et al.}~\cite{wright2026double} prepare the thermofield double on a doubled register by
running a double-bracket flow whose generator is the state-dependent commutator $[\Psi,H]$, with
$\Psi = |\psi\rangle\langle\psi|$ the projector onto the current state, building on
the double-bracket imaginary-time evolution of Ref.~\cite{gluza2026double}. The same state is the target of
the variational thermofield-double preparation of Ref.~\cite{wu2019variational}. Its resource statement
is a count of queries to $e^{-iHt}$ and to reflections about the current state. The paper contains no
gate counts, no circuit diagrams and no resource table.

The third family is local-driving Lindbladian simulation.
Hahn, Parameswaran and Placke~\cite{hahn2025towards} approximate an exact KMS-detailed-balanced Gibbs
sampler, of the family established in Ref.~\cite{chen2023quantum}, using only analog evolution
under $H$, strictly local time-dependent couplings to ancillas, and ancilla reset. Its resource
statement is a total Hamiltonian evolution time, written $T_{\mathrm{tot}} = O(N_{\mathrm{step}}T)$ in its
own resource estimate (v2 Sec.~IV~C).

These three sit inside a larger family. The exactly KMS-detailed-balanced Gibbs samplers of
Refs.~\cite{chen2023quantum,chen2023efficient,ding2025efficient} are the theoretical backdrop all three
approximate. Rapid mixing at high temperature is now proved for them~\cite{rouze2026optimal}, together
with the result that at inverse temperature polynomial in system size implementing the evolution is
BQP-complete~\cite{rouze2026efficient}. Around these sit Lindblad-engineering constructions under the
eigenstate thermalization hypothesis~\cite{brunner2025lindblad}, randomized Lindblad
simulation~\cite{chen2025randomized}, near-term system--bath variants~\cite{lloyd2025quantum}, Davies
generators with engineered rounding promises~\cite{rall2023thermal}, and the QITE
family~\cite{motta2020determining,tang2026interference}. Mixing time, the quantity all of these ultimately
depend on, is itself hard to certify~\cite{ramkumar2024mixing}. The exact samplers descend from quantum
Metropolis, phase-estimation, Szegedy-walk and Lindblad-simulation
algorithms~\cite{temme2011quantum,yung2012quantum,poulin2009sampling,chowdhury2017quantum,wocjan2023szegedy,cleve2017efficient},
and constant temperature already suffices for a provable advantage on suitably chosen
Hamiltonians~\cite{bergamaschi2024quantum}, though not on a 1D chain. Metropolis-type samplers, which
presuppose fault tolerance, and variational thermal-state
preparation~\cite{verdon2019quantum,chowdhury2020variational,sagastizabal2021variational,selisko2024extending},
whose accuracy is set by an ansatz, are not benchmarked here. Implementations that replace the block
encoding by dense local circuits~\cite{hahn2026efficient} or by analog evolution with local
drives~\cite{hahn2025towards} are not priced in gates and are likewise outside the comparison.

Three currencies (gate depth on one device, oracle queries, and analog evolution time) are not
comparable as published, and neither of the two documents most likely to contain a cross-method
comparison makes them so. A complete read of both (the 62-page Ref.~\cite{brunner2025lindblad} and the
supplement of Ref.~\cite{ding2025simple}) found no such comparison: the latter's only comparison table rates eleven
algorithms on three qualitative properties, a fixed-point error bound, a mixing-time guarantee and an
early-fault-tolerant implementation, with check marks, crosses and question marks, alongside a free-text
remarks column, and it carries no cost column of any kind. Table~II of Ref.~\cite{wright2026double} is a
comparison in a different currency, asymptotic query complexities with no constants, whose two cost columns
are populated for one row each out of eight (Sec.~\ref{sec:reimpl-db}).

Hahn, Parameswaran and Placke~\cite{hahn2025towards} close their
discussion by noting that ``\ldots it would be interesting to ask how these methods compare operationally
with others for controlled approximation of thermal observables\ldots''. Wright \emph{et
al.}~\cite{wright2026double} record the same open problem in their own discussion
(Sec.~\ref{sec:reimpl-db}). We address that question directly.

Idealized primitive counts are known to overstate quantum advantage once
compiled~\cite{babbush2021focus,hoefler2023disentangling,beverland2022assessing,dalzell2025quantum,sanders2020compilation,jaques2025qram}.
Ref.~\cite{liu2026logical} compiles competing state-preparation circuits to a common Clifford$+T$ set at
matched accuracy, but for one static primitive without dynamics or temperature. Rankings that invert once
every arm is tuned to the same standard, and the case for reporting tuning budgets, are documented in
quantum computing~\cite{bonetmonroig2023performance,proctor2025benchmarking,bowles2024better}
and in machine learning~\cite{lucic2018gans,melis2018state,ferraridacrema2019progress,dodge2019show,tang2020showing,bouthillier2021accounting}.

Existing harnesses hold the algorithm fixed and vary the
device~\cite{lubinski2023application,chatterjee2025comprehensive}, or count logical resources without an
accuracy axis~\cite{harrigan2024qualtran}. Ref.~\cite{sharma2024quantum} prices a single Gibbs-property
algorithm in gates rather than competing ones. The ground-state V-score~\cite{wu2024variational} has no
state-computable finite-temperature analogue: certified relaxations~\cite{fawzi2024certified} bound
observables classically, and a mirror-circuit diagnostic~\cite{granet2026adiabatic} serves one protocol,
leaving minimally entangled typical thermal states~\cite{white2009minimally,stoudenmire2010minimally} and
thermal pure quantum states~\cite{sugiura2012thermal} as the classical references.

We contribute a harness and the frontier it produces, not a ranking. The harness rests on five
fairness rules, each fixed before the arm that would test it was implemented. Rule R1 declares side
channels. The headline resource is the cumulative two-qubit-gate count, and every artifact additionally
carries \texttt{(qubits, ancillas, resets, measured Pauli strings)}, because a gate count hides all four
and the three arms differ in all four. Rule R2 sets a per-arm step-size policy, not a global $\delta t$.
Each arm declares a rule with one free parameter, derived from the model, before results. A global step
size would be unfair to all three: one arm's step is bounded by a Nyquist condition on a bath window,
another's by a variational inner loop, the third's by a filter width. Rule R3 fixes one tuning-budget
currency: a fixed number of calls to the exact reference (\texttt{GibbsReference.metrics}) per arm per
$\beta$, the one currency all three arms could spend, counted automatically by a metered wrapper. The
budget is a cap of 200 calls per $\beta$, and what each arm actually spent is reported
(Sec.~\ref{sec:tuning}). Rule R4 tunes on free energy and reports both metrics. All arms are tuned on the
relative free-energy error, a one-sided quantity vanishing only at $\rho_\beta$, and reported on both it
and trace distance. The head-to-head is matched in trace distance, which bounds the error of every bounded
observable at once and is therefore the most demanding choice, growing with $n$ at fixed local error. Its
free-energy counterpart is in Sec.~\ref{sec:sensitivity}. Rule R5 makes the compilation regime a reported
axis: two costings (CNOT ladder and native $R_{ZZ}$) from one circuit, both reported, fixed as an axis in
advance rather than chosen after the fact.

We added five further requirements, labeled N1 to N5 below, after the second arm and imposed them before
the third. The most consequential is N1: each arm declares its cost model (oracle lower bound versus
realizable circuit) before any result, and every run carries a required boolean
\texttt{circuit\_is\_the\_simulation}. N2 places the non-Pauli primitive catalog in
\texttt{accounting.PRIMITIVES}, shared across arms rather than added per arm, alongside the shared
\texttt{TrotterPolicy}, since ad-hoc per-arm addition is how a common gate set stops being common. N3
declares a cost tie-break at 5\,\% tolerance (Sec.~\ref{sec:tuning}). N4 requires the mixing-time
parameter $t^\star$, the normalized inverse temperature $\bt = \beta\lambda$ with $\lambda$ the spectral
half-width (both in Sec.~\ref{sec:reference}), and the Lindbladian clock $kJ^2$, with $k$ the number of
cycles and $J$ the drive coupling, to be reported per cell. N5 enforces per-arm simulator ceilings in code rather than in documentation.

We use the harness to establish four results. We show that the three arms have qualitatively different
accuracy floors (Sec.~\ref{sec:floors}) and that rankings invert with the accuracy target, the gate budget
and the cost model (Sec.~\ref{sec:inversions}). We find that the gap between an algorithm's own resource
currency and a realizable circuit depends on the method (Sec.~\ref{sec:oraclegap}). We surface, by
reimplementation, discrepancies in all three source papers (Sec.~\ref{sec:reimpl}).
Sec.~\ref{sec:heisenberg} carries the first two results onto a second model family.

Sec.~\ref{sec:harness} describes the harness, its cost models and its tuning budgets.
Sec.~\ref{sec:results} reports the accuracy floors, the ranking inversions, the oracle-to-realizable gap,
the scaling in $\bt$ and the Heisenberg cross-check. Sec.~\ref{sec:reimpl} lists the discrepancies
reimplementation surfaced, and Sec.~\ref{sec:discussion} sets out how far each finding extends beyond this
grid. The appendices give the protocol implementations and the deviations from the source papers.

\section{The benchmarking harness}
\label{sec:harness}

The harness is 4348 lines of pure \texttt{numpy}/\texttt{scipy} in \texttt{src/tbyp}, CPU only, plus 2191
lines of tests and 3820 of sweep, table and figure scripts, 10\,359 in total. The three arms are built on
three shared modules, \texttt{models.py}, \texttt{reference.py} and \texttt{accounting.py}. The remainder
of this section states what the three arms must agree on.

\subsection{Schedule, compilation, and the counted-equals-simulated guarantee}
\label{sec:accounting}

The central structural guarantee is that a protocol is expressed as an ordered list of
\texttt{(PauliTerm, time)} pairs (a \emph{schedule}), and two functions consume that same schedule.
\texttt{compile\_schedule} emits a circuit and meters it, and \texttt{schedule\_unitary} builds the unitary
actually applied to the state. A reported gate count therefore always corresponds exactly to the reported
accuracy: there is no idealized simulation running alongside an idealized count. Where an arm's cost
model deliberately breaks this (Sec.~\ref{sec:costmodels}) the run carries
\texttt{circuit\_is\_the\_simulation = False} and every table labels it a lower bound.

Arbitrary single-qubit gates are counted but are not the headline. CNOT is the only entangler,
and mid-circuit reset is counted separately at zero two-qubit gates, which is the accounting the
139-qubit experiment itself uses. A weight-$w$ Pauli rotation $e^{-i\theta P}$ compiles to basis change, CNOT ladder,
$R_Z(2\theta)$, ladder$^{-1}$, basis change$^{-1}$: exactly $2(w-1)$ two-qubit gates, so a $ZZ$ factor is
$2\,\mathrm{CNOT}+1\,R_Z$ and a weight-1 factor is free. Connectivity as requested, no SWAP routing,
the same allowance for every arm, so absolute counts are lower bounds on hardware. Trotter order is
recorded in every artifact and is second order (Strang) for all three arms here. Non-Pauli primitives live
in one shared catalog, \texttt{accounting.PRIMITIVES}, each with a cited two-qubit count, an ancilla
count, an exact logical action for the statevector checker, and an explicit flag for whether the native
regime touches it. There is exactly one here, the double-bracket arm's multi-controlled phase.

The counts are hand-checkable. For the system--bath arm on $n$ sites plus one ancilla, $n$ weight-2 terms
$\times$ 2 CNOTs $\times$ order $\times$ $M$ steps gives 48 / 72 / 96 two-qubit gates per reset cycle at
$n = 4/6/8$ for the hardware preset and 768 / 1024 at $n = 6/8$ for the classical preset. For the
Lindbladian arm,
\begin{equation}
\twoq/\text{cycle} \;=\; 4(n-1)\bigl[\,2 r_{\text{wind}} + (M-1) r_{\text{drift}}\bigr] \;+\; M\, n_B\, 2 w_A ,
\label{eq:ldcount}
\end{equation}
with $r_\cdot = \lceil |\tau| \|H\|_1 / \text{angle\_budget}\rceil$ the Trotter repetition counts of the
rewind and drift blocks, $M$ the number of drive pulses per cycle, $n_B = n$ the number of bath ancillas,
and $w_A$ the Pauli weight of the jump operator $A_a$ alone, so $w_A = 1$ for the source's
single-site jumps $A_i = Y_i$. The coupling that is actually compiled is $A_a \otimes X_{B_a}$, of weight
$w_A + 1$, and a weight-$w$ Pauli rotation costs $2(w-1)$ two-qubit gates, hence $2 w_A$ per pulse per
ancilla. At $n=4$, $\sigma=0.5$, $\beta = 0.2$ the three terms of Eq.~\eqref{eq:ldcount} are
$4\cdot3\cdot2\cdot11 = 264$, $4\cdot3\cdot23\cdot1 = 276$ and $24\cdot4\cdot2\cdot1 = 192$, totalling
$732$, which is what the meter reports. The repetition counts $r_{\text{wind}} = 11$ and
$r_{\text{drift}} = 1$ follow from $\|H\|_1 = 7$, $\text{angle\_budget} = 1$ and the two block durations
$(T-\delta t)/2 = 1.4375$ and $\delta t = 0.125$.

Following rule R5, the meter emits two costings from the same circuit:
\texttt{n\_2q} (CNOT ladder) and \texttt{n\_2q\_native\_pauli}, which re-prices each ladder pair as one
native $R_{ZZ}$, i.e.\ $w-1$ instead of $2(w-1)$ gates. Non-Pauli primitives are \emph{not} re-priced,
which is the entire mechanism by which the regime can reorder arms (Sec.~\ref{sec:inversions}).

\subsection{Per-arm cost models}
\label{sec:costmodels}

Each arm declares its cost models before results (requirement N1). Six models across three arms:

\begin{table}[h]
\centering\footnotesize
\begin{tabular}{@{}llll@{}}
\toprule
Arm & Model & What is charged & counted $\equiv$ simulated \\
\midrule
System--bath   & \texttt{hardware}  & the whole Trotterized cycle, from the schedule & yes \\
System--bath   & \texttt{classical} & the whole Trotterized cycle, from the schedule & yes \\
DB-TFD         & \texttt{oracle}    & one $e^{\pm i\sqrt{s}H}$ pair, one reflection per step & no (lower bound) \\
DB-TFD         & \texttt{pf}        & the realizable group-commutator circuit & yes \\
Local driving  & \texttt{oracle}    & Hamiltonian blocks only, drive and resets free & no (lower bound) \\
Local driving  & \texttt{circuit}   & Hamiltonian blocks and drive pulses, from schedule & yes \\
\bottomrule
\end{tabular}
\end{table}

The two \texttt{oracle} models are not conveniences: each is \emph{the source paper's own accounting},
priced by our compiler. Ref.~\cite{wright2026double} counts queries to Hamiltonian evolutions and
reflections. Ref.~\cite{hahn2025towards} counts exactly one resource, the total Hamiltonian evolution
time. Reporting only the compiled costs would misrepresent those papers' own accounting. Reporting only
the query costs would reproduce the incommensurability this work is intended to remove. We therefore
report both axes, and the gap between them is the sharpest result here (Sec.~\ref{sec:oraclegap}).

\subsection{Tuning budgets}
\label{sec:tuning}

Rule R3 fixes the currency, which is the number of calls to \texttt{GibbsReference.metrics} as counted
by \texttt{MeteredReference}, and caps it at 200 per $\beta$ per arm. DB-TFD spent 160 (40 candidates
$\times$ 4 calls) at $n = 4$ over $\text{angle\_budget} \in \{2,1,0.5,0.25\}$ $\times$ five root orderings
$\times$ two polynomial families, and chose $\text{angle\_budget} = 0.25$, \texttt{real\_desc}, Chebyshev
at every $\beta$. Local driving spent 180 (36 candidates $\times$ 5 calls) over
$\sigma \in \{0.5,0.75,1,1.5,2,3\}$ $\times$ $J \in \{0.25,0.5\}$ $\times$
$\text{angle\_budget} \in \{1,0.5,0.25\}$, choosing $(\sigma,J,\text{angle}) = (0.75,0.25,1)$,
$(1.5,0.5,1)$, $(2,0.5,0.5)$ at $\beta = 0.2/1/5$. The system--bath presets are the source papers'
own published hyperparameter sets, not tuned by us: 0 tuning calls against 160 and 180. A separate
one-parameter-at-a-time sensitivity study maps the surrounding landscape, but it selects nothing. In both
tuned arms the policy was selected at $n = 4$ and transferred unchanged to $n = 6, 8$. That transfer is a
declared choice, not a justified one, and Sec.~\ref{sec:floors} shows the Lindbladian arm's $\sigma^\star$
does move with $n$ at $\beta = 0.2$.

The system--bath arm is the untuned one, and the tuning asymmetry compounds with two simulator choices that act against it. The \texttt{classical}
preset is run at $\delta t = 0.25$ rather than its source's 0.1 (deviation Arm~1 D1): at $n = 4$,
$\beta = 1$ halving the step lowers its floor from 0.0494 to 0.0332, by a third. And its $\omega$ quadrature,
$n_\omega = 12$, is not converged at $\beta = 5$, where it makes that arm's accuracies about 8\,\%
pessimistic (D3). Both effects are one-signed against the system--bath arm, and its \texttt{classical}
floors are the quantities that decide the tight-target cells: every ``n/r'' in the SB-cl column of
Table~\ref{tab:threearm} at $\td \le 0.05$ and $0.02$ is therefore conditional on the step we ran rather
than a statement about the preset as published. Those entries therefore hold at $\delta t = 0.25$, and the
preset at its published $\delta t = 0.1$ is the one control outside this grid.

The tuning budget has a limited scope. R3's 200 calls per $\beta$ govern policy selection only. They do
not govern the production sweep, which is much larger and is selected with knowledge of the exact answer,
and the three arms do not sweep comparably large spaces. Both are properties of this study rather than of
the algorithms:

\begin{center}\small
\setlength{\tabcolsep}{4pt}
\begin{tabular}{@{}lrrrl@{}}
\toprule
arm & tuning calls & swept configs & reference evaluations & frontier selection \\
    & per $\beta$  & per cell      & per cell              &                    \\
\midrule
system--bath  & 0 (untuned) & 2 presets & 80--400 per preset & none: presets as published \\
DB-TFD        & 160 of 200  & 56        & 236                & min over (config, iteration) \\
local driving & 180 of 200  & 28 / 20--24 / 4--24 & 708--34\,356 & min over (config, cycle) \\
\bottomrule
\end{tabular}
\end{center}
\noindent (the Lindbladian configuration counts are $n = 4$ / $6$ / $8$)

\noindent Across the whole grid that is 2124 exact-reference evaluations for DB-TFD, 147\,932 for the
Lindbladian arm and 3990 for the two system--bath presets. The reported frontiers of the two tuned arms
are therefore envelopes over a swept family, selected with exact-answer knowledge, against two untuned
single configurations for the system--bath arm. That generosity is recorded as deviation Arm~2 D8 for
DB-TFD. It applies with more force to the Lindbladian arm, which sweeps the wider space and
wins the most cells, and which is charged up to 145 times more reference evaluations per cell than DB-TFD
wherever the simulation is cheap enough to allow it. The envelope is narrowest at $n = 8$, where the Kraus
build is the binding cost and the arm is charged 3 to 6 times DB-TFD's 236 rather than 145 times it, so at
the size where an envelope would help it most it is the least generously treated.
Sec.~\ref{sec:budgetfrontier} therefore reports a second, budget-respecting frontier alongside the envelope,
and Sec.~\ref{sec:sensitivity} reports how far each headline count moves between them.

For DB-TFD the free-energy
error is \emph{monotone} in the discretization, so a pure-accuracy objective always selects the finest
Trotter step: at the tuning grid's $k = 3$ it bought $4.0\,\%$ / $0.9\,\%$ / $4.5\,\%$ of free-energy error
at $\beta = 0.2/1/5$ for $1.29\times$ / $1.38\times$ / $1.57\times$ more gates. Requirement N3 therefore
declared a cost tie-break: among candidates within 5\,\% of the best, take the cheapest. Applied to
that grid it selects $\text{angle\_budget} = 2.0 / 1.0 / 2.0$ at $\beta = 0.2/1/5$, against the
accuracy-optimal $0.25$ the tuning recorded as \texttt{chosen}.

The production sweep was launched at $0.25$, so N3 was not in force for any number reported below. The
consequence is bounded and one-signed: DB-TFD's gate counts here are $1.29$--$1.57\times$ above what the
declared policy would have charged it, so every cell DB-TFD wins it would still win under N3, while the
cells it loses are not determined by this bound. Sec.~\ref{sec:sensitivity} carries that through to the headline counts.
The tie-break is separately a no-op for the Lindbladian arm, whose accuracy is non-monotone in its
principal parameter and so already has an interior optimum, and it is also a no-op for DB-TFD's
sub-stepping depth, where the finest setting is never within 5\,\% of any coarser one. The general rule
that follows is to declare a cost tie-break per arm exactly when that arm's accuracy is monotone in a
cost-increasing parameter, and then to verify that the production sweep runs at the policy the tie-break
selects.

\subsection{Exact reference, metrics, and benchmark grid}
\label{sec:reference}

$\rho_\beta = e^{-\beta H}/Z$ and $F(\beta) = -\beta^{-1}\log Z$ by dense diagonalization with the
ground-state energy shifted out before exponentiating. Supported to $n = 12$, where one complex
density matrix is 0.25 GiB (1 GiB at $n = 13$).

Three metrics: the trace distance $\td = \tfrac12\|\rho - \rho_\beta\|_1$ (the headline accuracy axis),
the squared Uhlmann fidelity, and the relative free-energy error $|F(\rho) - F(\beta)|/|F(\beta)|$ with
$F(\rho) = \mathrm{tr}(\rho H) - S(\rho)/\beta$, with $S(\rho)$ the von Neumann entropy. Since
$F(\rho) - F(\beta) = \beta^{-1}D(\rho\|\rho_\beta) \ge 0$, with $D$ the quantum relative entropy, the
free-energy metric is one-sided and vanishes only at $\rho_\beta$. We check that identity numerically,
which ties the two metrics together
rather than leaving them as two implementations that could both be wrong.

A fourth quantity carries the most weight in what follows. \texttt{effective\_beta} returns the inverse
temperature whose exact Gibbs state has the same energy as the prepared state, the same quantity
Ref.~\cite{farrell2026preparing} fits against quantum Monte Carlo. It separates ``this state is far from
$\rho_\beta$'' from ``this state is a thermal state at the wrong temperature'', a distinction a trace
distance cannot make and on which two of the findings below turn.

The benchmark grid is the 1D transverse-field Ising model at criticality,
$H = -J\sum Z_iZ_{i+1} - h\sum X_i$ with $h = J$, open chain, $n \in \{4,6,8\}$,
$\beta \in \{0.2,1,5\}/J$. All arms started from the maximally mixed state, so all three begin at identical
accuracy. Targets $\td \le \{0.3, 0.1, 0.05, 0.02\}$, giving 36 matched $(n,\beta,\text{target})$ cells,
each evaluated under two compilation regimes and six cost models. A Heisenberg cross-check on a subgrid
is reported separately in Sec.~\ref{sec:heisenberg}.

The upper end of the size range $n \in \{4,6,8\}$ is a property of
the simulator rather than of the protocols. The exact reference reaches $n = 12$, and the system--bath and
double-bracket arms reach $n \approx 10$ and $n = 12$ in the same memory. The binding constraint is the
Lindbladian arm, whose one-ancilla-per-jump-operator design places a $16\cdot 8^{n_S}$-byte Kraus block in
memory, 256 MiB at $n = 8$, 2 GiB at $n = 9$ and 16 GiB at $n = 10$, and peak usage is about three times
the block (about 6 GiB at $n = 9$) because
the rotation kernel holds two temporaries of the same size. Requirement N5 enforces the resulting per-arm ceilings
in code rather than in documentation, so an absent $n = 10$ Lindbladian point records a simulator ceiling
and not a limit of the method, whose circuits are unremarkable at that size. The same
ceiling binds the mixing-time parameter $t^\star$ of Ref.~\cite{hahn2025towards}, which requires an ODE on
a $4^n$ state and is therefore computed at $n \le 6$ only.

Every complexity theorem in this literature is stated for $\|H\|_{\mathrm{op}} \le 1$, so the quantity those
theorems are about is not the physical $\beta$ but the \emph{normalized} inverse temperature
$\bt = \beta\lambda$ with $\lambda = (E_{\max} - E_{\min})/2$ (requirement N4). For 1D TFIM at $h = J$,
$\lambda = 4.759 / 7.296 / 9.838$ at $n = 4/6/8$, linear in $n$, so the grid spans

\begin{center}\small
\begin{tabular}{@{}lccc@{}}
\toprule
$\bt = \beta\lambda$ & $\beta = 0.2$ & $\beta = 1$ & $\beta = 5$ \\
\midrule
$n = 4$ ($\lambda = 4.759$) & 0.95 & 4.76 & 23.79 \\
$n = 6$ ($\lambda = 7.296$) & 1.46 & 7.30 & 36.48 \\
$n = 8$ ($\lambda = 9.838$) & 1.97 & 9.84 & 49.19 \\
\bottomrule
\end{tabular}
\end{center}

\noindent a factor of 51.7 in $\bt$ from a factor of 25 in $\beta$. Every table below is read on the
$\bt$ axis and every fit below is taken against it. Physical $\beta$ is retained as the label of the
experiment, because it is what an experimenter programs and what the source papers quote. Both are printed
in every table and carried in every artifact as \texttt{beta\_normalised}. Sec.~\ref{sec:clock} shows what
reading the same data on the physical axis costs.

\section{Results}
\label{sec:results}

\begin{figure}[t]
\centering
\includegraphics[width=0.95\textwidth]{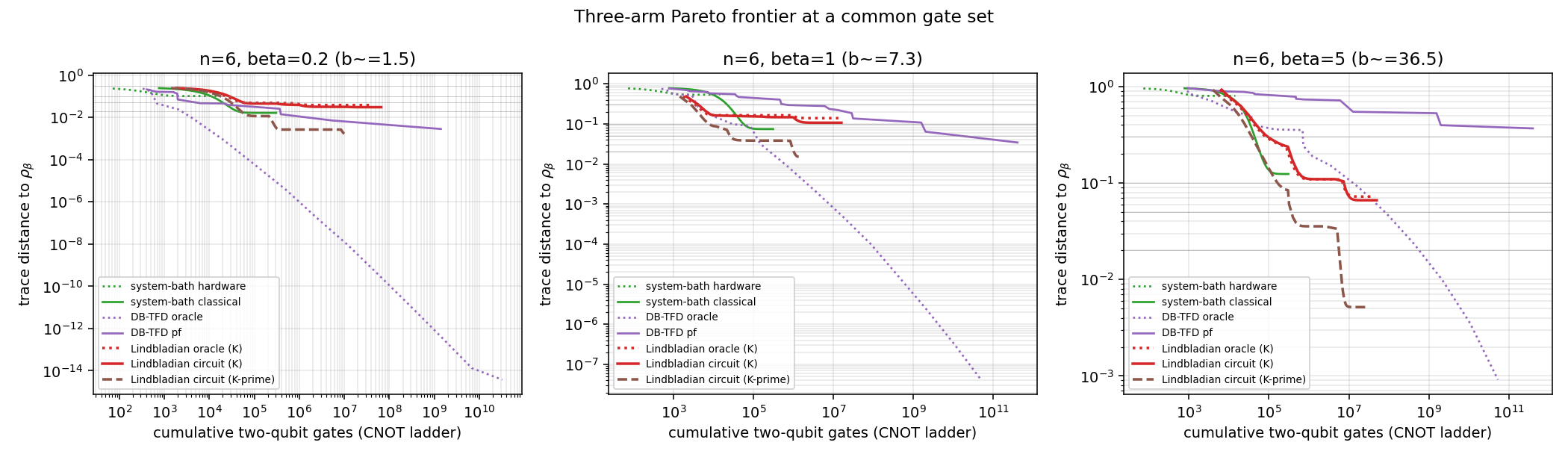}
\caption{The three-arm Pareto frontier at a common gate set, TFIM $n = 6$, $h = J$. Trace distance to the
exact Gibbs state against cumulative two-qubit gates in the CNOT-ladder regime, at
$\beta = 0.2/1/5$ ($\bt = \beta\lambda = 1.5/7.3/36.5$). Seven curves: two system--bath presets, DB-TFD
under its own query accounting and as a realizable product-formula circuit, and the local-driving arm
under both cost models and both channel variants. The three arms fail in three different ways: the
sampler flattens onto a bias floor, DB-TFD's oracle curve keeps falling to machine precision while its
compiled curve stalls, and the Lindbladian curves floor at a level set by a tunable parameter. That is why they
cross. The $n = 4$ and $n = 8$ panels behave the same way. The $n = 8$ Lindbladian grid matches the
$n = 6$ grid in $\sigma$ at $\beta = 0.2$ and is narrower at the other two temperatures (two widths at
$\beta = 5$, one at $\beta = 1$, both cost models and both channels throughout) because $n_S = 8$ is that
arm's exact-simulation ceiling, where one configuration costs 20--60 minutes end to end and the width
sweep at one temperature costs several hours. The system--bath \texttt{hardware} curves here are at that preset's
stated constant envelope, $\sigma = \infty$ (Sec.~\ref{sec:reimpl-sb}).}
\label{fig:frontier6}
\end{figure}

Figure~\ref{fig:frontier6} is the principal result. It shows not a winner but a structure: three arms
whose cost--accuracy curves have three different geometries and therefore cross.

\subsection{Three accuracy-floor geometries}
\label{sec:floors}

\begin{figure}[t]
\centering
\begin{minipage}[t]{0.36\textwidth}
\centering
\includegraphics[width=\textwidth]{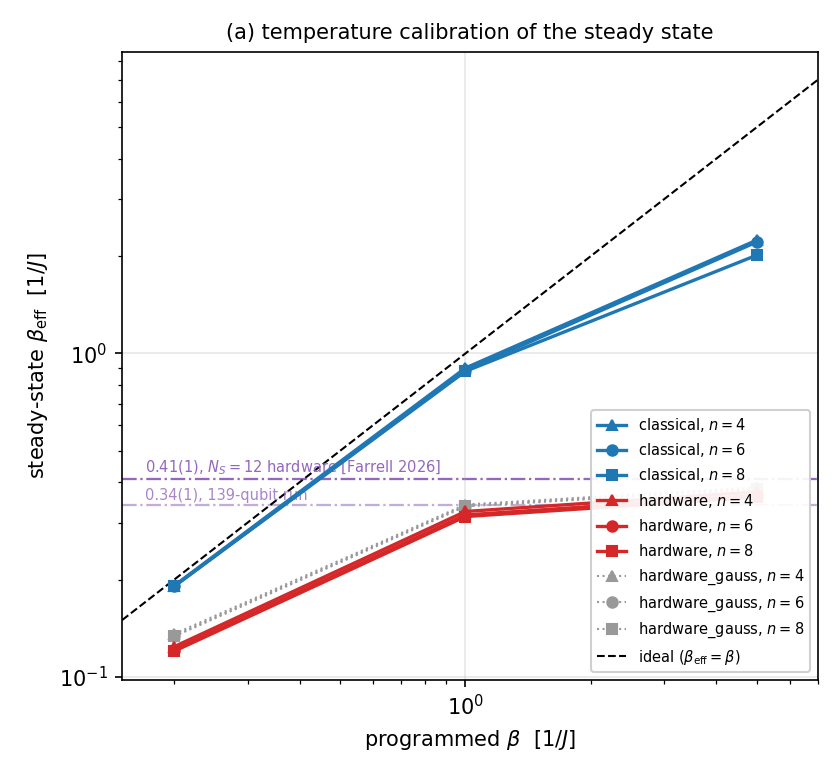}
\end{minipage}\hfill
\begin{minipage}[t]{0.60\textwidth}
\centering
\includegraphics[width=\textwidth]{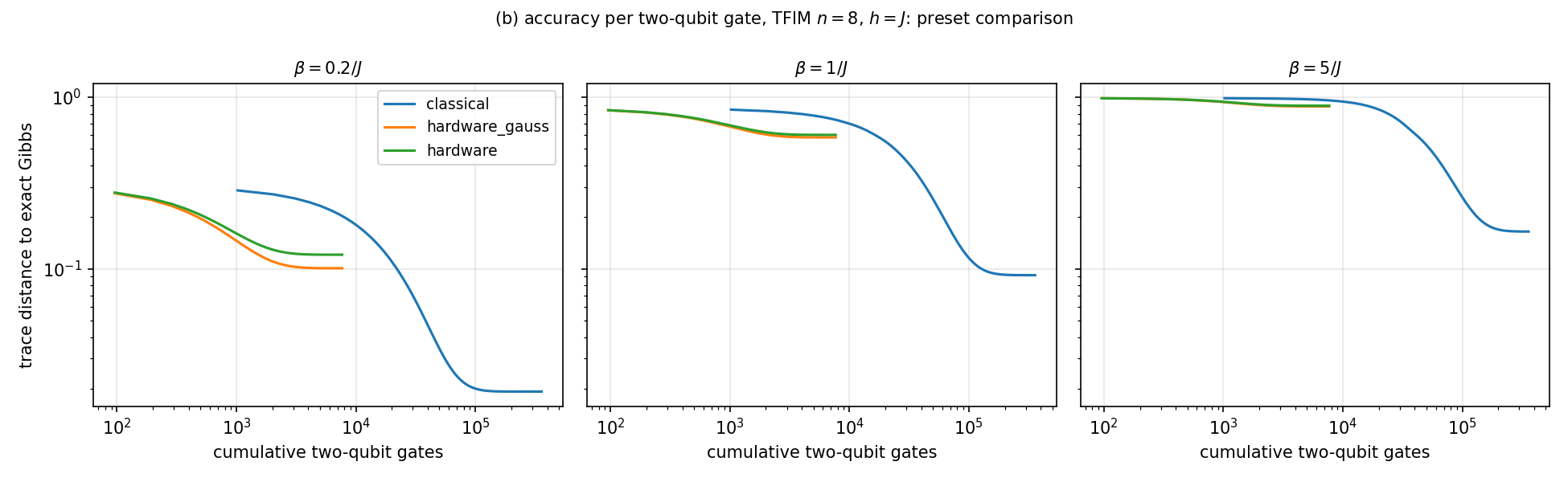}
\end{minipage}
\caption{System--bath temperature ceiling and preset crossover. (a) Steady-state $\betaeff$ against the
programmed $\beta$. The \texttt{hardware} preset, at the constant envelope its source specifies, saturates
below $0.4/J$ and the three sizes lie almost on top of each other. (b) The crossover \emph{within}
one arm, at $n = 8$. The cheap preset is $10.7\times$ ahead at $\td \le 0.3$ (96 gates against 1024) and
then flattens at 0.1207, missing $\td \le 0.1$ altogether, while the expensive preset clears it at
22\,528 and reaches 0.05 at 38\,912.}
\label{fig:betaeff}
\label{fig:presets}
\end{figure}

The first geometry belongs to the system--bath arm. At the 139-qubit experiment's hyperparameters, \emph{including the constant coupling envelope
$\sigma = \infty$ its Methods~D specifies}, the channel reaches a genuine fixed point within 14--35 reset
cycles and then stops: running 200 cycles instead of 40 changes the trace distance in the fourth decimal,
and the last-decile drift is 0.00\,\% to printed precision at every cell. Where it stops is a
thermal-\emph{looking} state at the wrong temperature (Fig.~\ref{fig:betaeff}, Table~\ref{tab:sb}):
$\betaeff = 0.124 / 0.325 / 0.373$ at $n = 4$, $0.121 / 0.317 / 0.364$ at $n = 6$ and
$0.120 / 0.313 / 0.359$ at $n = 8$ for requested $\beta = 0.2/1/5$. The effective inverse temperature
saturates below $0.4/J$, so that requesting $\beta = 5$ yields little more than requesting $\beta = 1$,
and the ceiling is size-independent to about 4\,\% over $n = 4 \to 8$, drifting slightly downward with $n$
rather than being constant to three digits, so it is a property of the
$(\alpha, T, \sigma, \omega_{\max})$ tuple and not of the model.

Ref.~\cite{farrell2026preparing} reports the same ceiling on hardware, at the preset it specifies:
``On the hardware side, the lowest effective temperature
we reach is $\betaeff = 0.41(1)$, constrained by the number of Trotter steps that can be executed in a
reset cycle without significant noise accumulation.'' That $0.41(1)$ is its $N_S = 12$ point. Its
139-qubit run reaches $0.34(1)$. Our exactly-simulated, noiseless $0.359$--$0.373$ at $\beta = 5$ falls
between the two, which locates the effect where that sentence locates it, in the reset-cycle budget, and
shows that it is not a noise effect: our simulation has no noise at all. The paper's own practice of
fitting a $\betaeff$ against quantum Monte Carlo, rather than claiming the $\beta$ it programmed, is the
same reading. Ref.~\cite{farrell2026preparing} chose that tuple under the hardware-noise constraint the
quoted sentence states and reports a fitted $\betaeff$ rather than the programmed $\beta$, so the ceiling
is a disclosed property of the preset, not a discrepancy with its results.

The same code at that paper's own \emph{classical-benchmark} settings ($T = 8$, $\sigma^2 = 1/16$,
$\alpha = 1$, $\omega_{\max} = 8$, all inside its documented range) converges properly, to
$\td = 0.0164 / 0.0742 / 0.1239$ at $n = 6$ with relative free-energy errors
$1.8\times10^{-4}$ / $2.5\times10^{-3}$ / $1.0\times10^{-2}$. The two presets are two regimes, not two
points on one frontier: at $n = 6$, $\beta = 0.2$ the hardware preset reaches $\td \le 0.3$ after 72
cumulative two-qubit gates against the classical preset's 768 ($10.7\times$ cheaper), then flattens at
0.1029 and never reaches 0.1, where the classical preset gets there at 11\,520 and to 0.05 at
20\,736 (Fig.~\ref{fig:presets}). A single-configuration comparison against either other arm
therefore comes out either way depending on the target.

\begin{table}[t]
\centering\small
\caption{System--bath arm: fixed points and their temperatures. Writing $d_j$ for the trace distance after
$j$ recorded points and $N$ for their number, \texttt{drift} is $|d_N - d_{N-k}|/d_N$ with
$k = \lceil N/10\rceil$, the one convention used for every drift reported here, so
\texttt{drift}$\,\approx 0$ certifies a fixed point rather than a truncated run. \texttt{hardware} is the
139-qubit run's hyperparameters \emph{with its stated constant envelope}, $\sigma = \infty$
(Sec.~\ref{sec:reimpl-sb}). \texttt{hardware\_gauss} is the same hyperparameters with the classical
benchmark's Gaussian $\sigma^2 = 1/16$ imported on top, which is not what that paper specifies and is shown
only to establish that the envelope shape is not what sets this preset's fixed point. Every row is a
converged fixed point: the largest drift in the table is 0.20\,\%.}
\label{tab:sb}
\begin{tabular}{@{}llrrrrrrrr@{}}
\toprule
preset & $n$ & $\beta$ & $\bt$ & cycles & \twoq/cyc & $\td$ & $\Delta F/F$ & $\betaeff$ & drift \\
\midrule
hardware        & 4 & 0.2 &  0.95 & 100 &   48 & 0.0792 & 0.00664 & 0.1241 & 0.00\,\% \\
hardware        & 4 & 1 &  4.76 & 100 &   48 & 0.4467 & 0.15515 & 0.3248 & 0.00\,\% \\
hardware        & 4 & 5 & 23.79 & 100 &   48 & 0.7288 & 0.41174 & 0.3733 & 0.00\,\% \\
hardware        & 6 & 0.2 &  1.46 & 200 &   72 & 0.1029 & 0.00739 & 0.1214 & 0.00\,\% \\
hardware        & 6 & 1 &  7.30 & 200 &   72 & 0.5339 & 0.16167 & 0.3169 & 0.00\,\% \\
hardware        & 6 & 5 & 36.48 & 200 &   72 & 0.8003 & 0.41433 & 0.3638 & 0.00\,\% \\
hardware        & 8 & 0.2 &  1.97 & 80 &   96 & 0.1207 & 0.00777 & 0.1202 & 0.00\,\% \\
hardware        & 8 & 1 &  9.84 & 80 &   96 & 0.6024 & 0.16483 & 0.3133 & 0.00\,\% \\
hardware        & 8 & 5 & 49.19 & 80 &   96 & 0.8910 & 0.41589 & 0.3594 & 0.00\,\% \\
\midrule
classical       & 4 & 0.2 &  0.95 & 200 &  512 & 0.0130 & 0.00016 & 0.1915 & 0.00\,\% \\
classical       & 4 & 1 &  4.76 & 200 &  512 & 0.0494 & 0.00199 & 0.9015 & 0.02\,\% \\
classical       & 4 & 5 & 23.79 & 200 &  512 & 0.1135 & 0.01138 & 2.2387 & 0.20\,\% \\
classical       & 6 & 0.2 &  1.46 & 400 &  768 & 0.0164 & 0.00018 & 0.1908 & 0.00\,\% \\
classical       & 6 & 1 &  7.30 & 400 &  768 & 0.0742 & 0.00245 & 0.8876 & 0.00\,\% \\
classical       & 6 & 5 & 36.48 & 400 &  768 & 0.1239 & 0.01033 & 2.2102 & 0.00\,\% \\
classical       & 8 & 0.2 &  1.97 & 350 & 1024 & 0.0192 & 0.00019 & 0.1907 & 0.00\,\% \\
classical       & 8 & 1 &  9.84 & 350 & 1024 & 0.0917 & 0.00271 & 0.8807 & 0.00\,\% \\
classical       & 8 & 5 & 49.19 & 350 & 1024 & 0.1647 & 0.01258 & 2.0118 & 0.04\,\% \\
\midrule
\multicolumn{10}{@{}l}{\emph{envelope-shape control (Gaussian envelope, not the published preset):}}\\
hardware\_gauss & 6 & 0.2 &  1.46 & 200 &   72 & 0.0862 & 0.00517 & 0.1342 & 0.00\,\% \\
hardware\_gauss & 6 & 1 &  7.30 & 200 &   72 & 0.5163 & 0.14714 & 0.3381 & 0.00\,\% \\
hardware\_gauss & 6 & 5 & 36.48 & 200 &   72 & 0.7955 & 0.39644 & 0.3822 & 0.00\,\% \\
\bottomrule
\end{tabular}
\end{table}

The second geometry belongs to the double-bracket arm. DB-TFD's error falls with the iteration count $k$ at every cell we ran, monotonically at eight of the nine
and with one exception at $n = 4$, $\beta = 5$, where $k = 8$ reaches $4.8\times10^{-4}$ and $k = 9$
returns $1.1\times10^{-3}$ before the descent resumes. In the \texttt{oracle} model the grid ceiling
$k = 12$ reaches $\td = 1.0\times10^{-15}$ at $n = 4$, $\beta = 0.2$, $3.8\times10^{-15}$ at $n = 6$,
$\beta = 0.2$ and $2.1\times10^{-10}$ at $n = 4$, $\beta = 1$ (Fig.~\ref{fig:dbtfd}). The full set is the
DB-TFD column of Table~\ref{tab:floors}. DB-TFD's limit is budget, not bias. The realizable product-formula
circuit converges to the \emph{same} states (sub-stepping is an exact splitting) at a very different
price (Sec.~\ref{sec:oraclegap}).

The third geometry belongs to the Lindbladian arm. This arm calibrates correctly where the system--bath arm does not. At the filter width its \emph{tuning}
selected, the $K'$ channel's fixed point sits at $\betaeff = 0.2012 / 1.0260 / 4.4698$ for requested
$\beta = 0.2/1/5$ at $n = 4$. At the width that minimizes the floor it reaches
$\betaeff = 0.2011 / 1.0260 / 5.0409$, the third of those at $\sigma = 3$ rather than the tuned $\sigma = 2$.
Both sets are given because the difference at $\beta = 5$ is the largest single effect of post-hoc
selection anywhere in this work (Sec.~\ref{sec:budgetfrontier}).

Its floor is set by the filter width $\sigma$, and $\sigma$ has an interior optimum that moves with
$\beta$: taking $\sigma^\star$ to be the swept width minimizing a channel's final trace distance, the
measured values are

\begin{center}\small
\begin{tabular}{@{}lccc@{}}
\toprule
$\sigma^\star$ & $\beta = 0.2$ & $\beta = 1$ & $\beta = 5$ \\
\midrule
$n = 4$, $K$  & 0.75 & 2.0 & 3.0 \\
$n = 4$, $K'$ & 1.0  & 1.5 & 3.0 \\
$n = 6$, $K$  & 1.5  & 1.5 & 3.0 \\
$n = 6$, $K'$ & 1.5  & 1.5 & 3.0 \\
$n = 8$, $K$  & $1.0^\S$ & $(1.5)^\dagger$ & $3.0^\ddagger$ \\
$n = 8$, $K'$ & $0.75^\S$ & $(1.5)^\dagger$ & $3.0^\ddagger$ \\
\bottomrule
\end{tabular}
\end{center}

\noindent The swept widths behind each entry are $\{0.5, 0.75, 1, 1.5, 2, 3, 4\}$ at $n = 4$.
At $n = 6$ they are $\{0.5, 0.75, 1, 1.5, 2, 3\}$ at $\beta = 0.2, 1$ and $\{1, 1.5, 2, 3, 4\}$ at
$\beta = 5$. At $n = 8$ they are the full $n = 6$ grid $\{0.5, 0.75, 1, 1.5, 2, 3\}$ at $\beta = 0.2$
($\S$), the single width the $n = 4$ tuning chose at $\beta = 1$ ($\dagger$), and the pair $\{2, 3\}$ at
$\beta = 5$ ($\ddagger$), because $n_S = 8$ is that arm's exact-simulation ceiling. A $\dagger$
entry is the transplanted width and not a measured argmin: agreement with the value that was transplanted
carries no information, and nothing below is inferred from those two cells.

The $\S$ entries are measured argmins over the same six widths as $n = 6$, but at a 300-cycle horizon
rather than the 12\,000 cycles used at $n = 4, 6$, because at $n_S = 8$ every point is an exact Kraus
build. The effect of the horizon is measurable from the existing artifacts:
truncating the $n = 4$ and $n = 6$ runs to 300 cycles moves $\sigma^\star$ from $0.75/1$ to $0.75/0.75$
and from $1.5/1.5$ to $1/0.75$ for $K/K'$, i.e.\ it moves the argmin \emph{left}. At a matched 300-cycle
horizon the three sizes give $0.75/0.75$, $1/0.75$ and $1/0.75$, so $n = 6$ and $n = 8$ agree in both
channels and $n = 4$ is one grid step below them in $K$.

One conclusion at $n = 8$ does not depend on the horizon at all. At $\beta = 0.2$ the $\sigma = 0.5$ and
$\sigma = 0.75$ runs have \emph{converged} by cycle 300 (last-decile drift 0.01\,\% and 0.21\,\% in $K$,
0.00\,\% and 0.37\,\% in $K'$), while every wider width is still descending monotonically. Since $K$ at
$\sigma = 1$ already reads 0.0391 against $\sigma = 0.75$'s converged 0.0445 and $\sigma = 0.5$'s
converged 0.0625, and can only fall further, the rewound channel's $\sigma^\star$ at $n = 8$ is at
least 1 at any horizon. The width the $n = 4$ tuning transplanted is therefore not the argmin at $n = 8$,
and it is wrong in the same direction as at $n = 6$. For $K'$ the same argument bounds $\sigma^\star$
below by 0.75 but no further, because $\sigma = 1$ is still falling at 29\,\% per decile when the run
stops. Which of the two wins at convergence is not settled here.

The optimum therefore moves right with $\beta$ at all three sizes, and it is not size-independent.
Comparing $n = 4$ with $n = 6$ at their common 12\,000-cycle horizon, it agrees at $\beta = 5$, where both
channels select $\sigma = 3$ against grids that reach 4, and at $\beta = 1$ for $K'$. It disagrees by
one or two grid steps at $\beta = 0.2$ for both channels and at $\beta = 1$ for $K$, where $n = 4$ selects
2.0 against $n = 6$'s 1.5. The $n = 8$ sweep extends the $\beta = 0.2$ statement to a third size in the
same direction: $\sigma^\star \ge 1 > 0.75$ for $K$ there, against $n = 4$'s 0.75. What the $n = 8$
horizon cannot decide is whether the rightward movement continues past $n = 6$ or has flattened, since the
matched-horizon argmins agree at those two sizes while the converged ones are available at only one of
them. At $\beta = 0.2$ the curve is
also bimodal in $K'$, with a local minimum at $\sigma = 0.75$ and the global one further right: at $n = 4$
the floors are $0.0029 / 0.0014 / 0.0016$ at $\sigma = 0.75/1/1.5$ and at $n = 6$ they are
$0.0032 / 0.0040 / 0.0016$, so the curves have the same shape and disagree about which lobe wins.
Size-independence of $\sigma^\star$ is therefore not available as a justification for the
$n = 4 \to 6,8$ transfer of Sec.~\ref{sec:tuning}. The transfer is a declared policy choice whose cost is
measured in Sec.~\ref{sec:budgetfrontier} instead.

The two failure modes on either side of the optimum are visibly different, and the distinction is a
statement about the fixed point rather than about the horizon, because every run quoted in this paragraph
is 12\,000 cycles.
Below $\sigma^\star$ the state is thermal at the wrong temperature ($\betaeff$ too small, drift zero: a
bias). Above it the state stops arriving: at $n = 4$, $\beta = 0.2$ the $K'$ channel gives
$\betaeff = 0.2011 / 0.2010 / 0.1984 / 0.0400 / 0.0111$ at $\sigma = 1 / 1.5 / 2 / 3 / 4$, so the calibration
survives one grid step past the optimum at a mixing time that is already too long to certify (11.3\,\%
drift at $\sigma = 2$ after 12\,000 cycles) and then collapses. The mechanism is measurable in the paper's own
currency, its mixing-time parameter $t^\star$ computed on the \emph{exact} sampler. At $n = 4$,
$\beta = 0.2$ that is $1.35 \to 4.18 \to 10.7 \to 70.8$ across $\sigma = 0.5, 0.75, 1, 1.5$, with no finite
estimate from $\sigma = 2$ up. Because the $\sigma$ grid is not uniform the growth must be quoted per step
rather than per unit: the two $\Delta\sigma = 0.25$ steps give $3.10\times$ and $2.56\times$ and the
$\Delta\sigma = 0.5$ step gives $6.61\times$. At $\beta = 1$ the four steps give $1.72$, $3.12$, $7.72$ and
$5.44\times$. At $\beta = 5$ the sequence is \emph{not} monotone, $t^\star = 1.55, 1.68, 1.45, 9.57, 271$
at $\sigma = 0.75, 1, 1.5, 2, 3$, which is the same non-monotonicity that puts $\sigma^\star$ at 3 there.
This is the prediction of Ref.~\cite{hahn2025towards}'s own Sec.~IV~D (``Optimal Accuracy'', present in v2
only), here measured in gates.

\begin{figure}[t]
\centering
\includegraphics[width=0.82\textwidth]{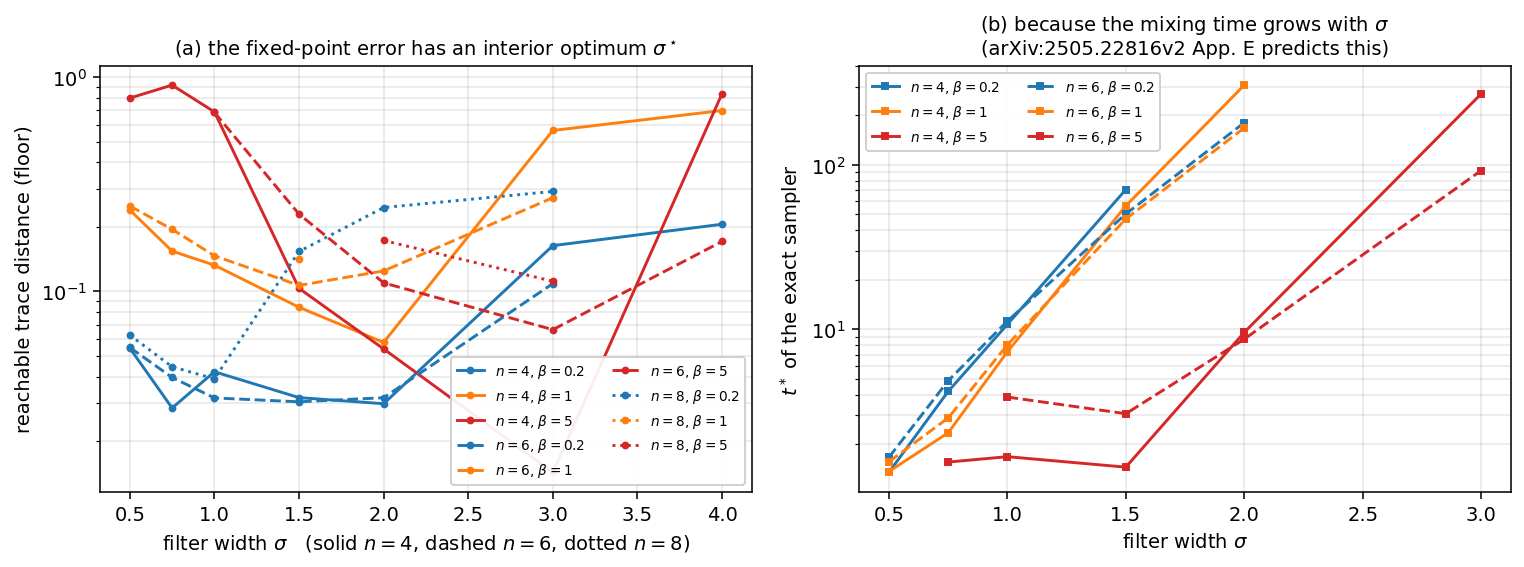}
\caption{Lindbladian floor and mixing time against filter width $\sigma$. (a) Reachable trace distance (the
final value, as in Table~\ref{tab:floors}) against filter width $\sigma$, for the rewound channel $K$:
non-monotone with an interior minimum at every $(n,\beta)$, and the minimum moves right as $\beta$ grows,
but not by the same amount at $n = 4$ and $n = 6$ (Sec.~\ref{sec:floors}). (b) The mechanism:
$t^\star$, the source paper's own mixing-time parameter evaluated on the exact KMS-detailed-balanced
sampler, rising by $1.7$--$7.7\times$ per grid step at $\beta = 0.2, 1$ and non-monotonically at
$\beta = 5$. The $\sigma$ grid is not uniform, so the growth is quoted per step and not per unit.
Points with no finite $t^\star$ estimate within the integration horizon are absent rather than plotted at
the horizon. Solid/dashed distinguish $n = 4$ from $n = 6, 8$. The $n = 8$ curve in (a) is a full six-width
sweep at $\beta = 0.2$, taken at that size's 300-cycle horizon, and is two points at $\beta = 5$ and one at
$\beta = 1$, and $t^\star$ is computed at $n \le 6$ only because it needs an ODE on a $4^n$ state.}
\label{fig:sigma}
\end{figure}

\begin{figure}[t]
\centering
\includegraphics[width=0.95\textwidth]{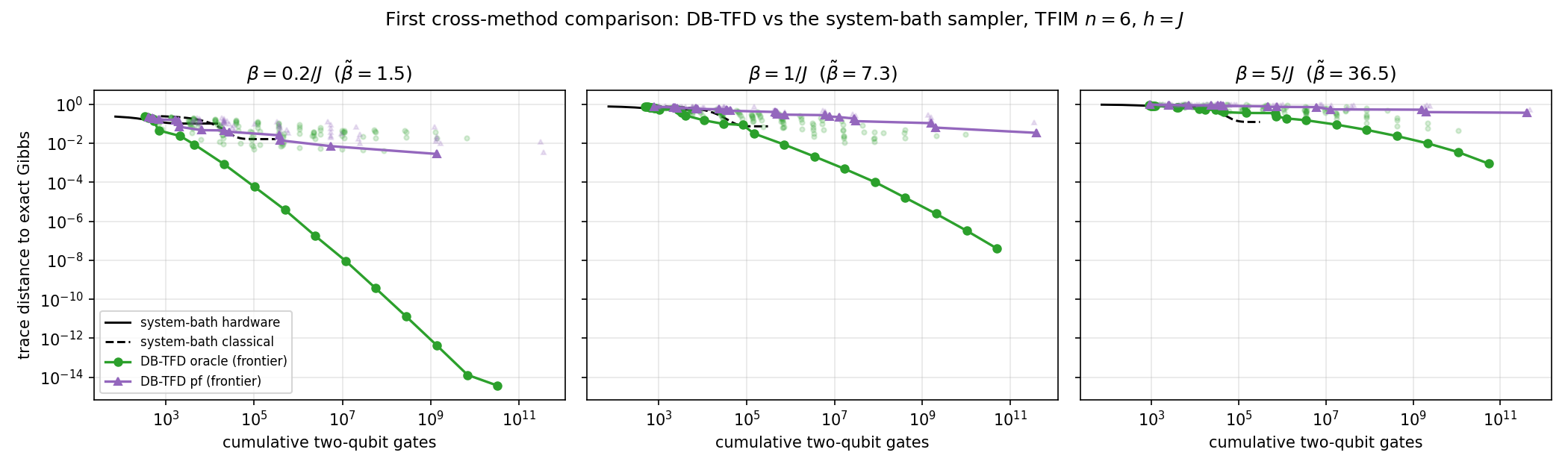}
\caption{The oracle/circuit split at $n = 6$. The two DB-TFD curves are the \emph{same algorithm} at the
same accuracies, priced under its own query accounting (green) and as a realizable group-commutator
circuit (purple). Faint markers are the individual swept configurations behind each frontier. The
system--bath curves flatten onto their bias floors.}
\label{fig:dbtfd}
\end{figure}

That divergence also bounds what the last-decile drift test can certify. The test separates a fixed point
from a truncated trajectory only while the mixing time is small against the horizon, and above
$\sigma^\star$ the mixing time is the quantity that has diverged: $t^\star$ is already 168--306 at
$\sigma = 2$ in the three cells where $\sigma = 2$ exceeds $\sigma^\star$, and it has no finite estimate
within the integration horizon at $\sigma = 3$ in four of the six cells where it is computed at all. At
$\beta = 5$, where $\sigma^\star$ is 3 and $\sigma = 2$ therefore sits below it, the same quantity is
8.7--9.6 and the test is informative. The horizons were chosen accordingly, so that every Lindbladian
entry reported in Tables~\ref{tab:floors} and \ref{tab:krewind} carries a drift at or below 1\,\% apart
from the single exception recorded in those captions: seven of the nine no-rewinding rows are at 0.00\,\%,
and the largest drifts in the selected set are the $n = 8$, $\beta = 5$ entries at 0.57\,\% and 1.00\,\%,
which are the largest $\bt$ in the grid at the widest selected $\sigma$ and therefore where
Sec.~IV~D of Ref.~\cite{hahn2025towards} places the worst mixing time. Wide-$\sigma$ points that are not
selected carry larger drifts. They appear in Fig.~\ref{fig:sigma}(a) and enter no table.

Table~\ref{tab:floors} collects the three geometries. At its own best swept width the Lindbladian $K'$
channel's floor is below the system--bath \texttt{classical} floor at \emph{every} cell, by factors of
$9.1$, $2.8$, $47.8$, $10.4$, $4.7$, $24.0$, $2.3$, $1.7$ and $10.5$ reading down the table, i.e.\
$1.7$--$48\times$. The advantage is not systematically largest at $\beta = 5$: it is $47.8$, $24.0$ and
$10.5$ there against $9.1$, $10.4$ and $2.3$ at $\beta = 0.2$, so it tracks $n$ at least as strongly as it
tracks $\beta$, and its smallest value falls at the largest $\bt$ of the $\beta = 1$ column rather than
anywhere in the $\beta = 0.2$ one.

\begin{table}[t]
\centering\small
\caption{Best fixed point reached by each arm at matched $(n,\beta)$, under one declared convention:
the final recorded trace distance of the best configuration in that arm's swept grid, never a
transient minimum. $K$ is the source paper's rewound protocol and $K'$ its own no-rewinding variant
(Sec.~\ref{sec:reimpl-ld}), and $\sigma^\star$ is the swept filter width minimizing that channel's final trace
distance, chosen per row and per channel, and is printed because it differs between the two channels at two
cells. The $\sigma$ grids include $\sigma = 3$ at every $\beta = 5$ cell, $n = 8$ included, so the
$\beta = 5$ rows are like-for-like across sizes. At $n = 8$, $\beta = 1$ the swept grid holds one width, so
that $\sigma^\star$ entry records the width transplanted from the $n = 4$ tuning rather than a measured
argmin, and the floor beside it is the floor at that width (Sec.~\ref{sec:floors}). DB-TFD is its
\texttt{oracle} model at the grid ceiling
$k = 12$, so every DB entry is budget-limited rather than floored. \texttt{drift} is the last-decile
convention of Table~\ref{tab:sb}. Every entry is at or below 1\,\% except the $n = 8$,
$\beta = 0.2$ rewound one at 10.41\,\%, whose run is still descending monotonically at the 300-cycle
horizon that size allows, so that single floor is an upper bound on its fixed point rather than the fixed
point (Sec.~\ref{sec:floors}). Every other trace distance here is a converged value and not a truncated
trajectory.}
\label{tab:floors}
\setlength{\tabcolsep}{4pt}
\begin{tabular}{@{}lllrrlrrrrrr@{}}
\toprule
& & & \multicolumn{2}{c}{SB \texttt{classical}} & DB-TFD & \multicolumn{3}{c}{Lindbladian $K$}
& \multicolumn{3}{c}{Lindbladian $K'$} \\
\cmidrule(lr){4-5}\cmidrule(lr){6-6}\cmidrule(lr){7-9}\cmidrule(lr){10-12}
$n$ & $\beta$ & $\bt$ & floor & drift & oracle, $k=12$ & $\sigma^\star$ & floor & drift & $\sigma^\star$ & floor & drift \\
\midrule
4 & 0.2 &  0.95 & 0.0130 & 0.00\,\% & $1.0\times10^{-15}$ & 0.75 & 0.0286 & 0.00\,\% & 1 & \textbf{0.0014} & 0.00\,\% \\
4 & 1 &  4.76 & 0.0494 & 0.02\,\% & $2.1\times10^{-10}$ & 2 & 0.0578 & 0.00\,\% & 1.5 & \textbf{0.0176} & 0.00\,\% \\
4 & 5 & 23.79 & 0.1135 & 0.20\,\% & $7.4\times10^{-5}$ & 3 & 0.0143 & 0.00\,\% & 3 & \textbf{0.0024} & 0.00\,\% \\
6 & 0.2 &  1.46 & 0.0164 & 0.00\,\% & $3.8\times10^{-15}$ & 1.5 & 0.0305 & 0.00\,\% & 1.5 & \textbf{0.0016} & 0.00\,\% \\
6 & 1 &  7.30 & 0.0742 & 0.00\,\% & $4.2\times10^{-8}$ & 1.5 & 0.1067 & 0.00\,\% & 1.5 & \textbf{0.0158} & 0.00\,\% \\
6 & 5 & 36.48 & 0.1239 & 0.00\,\% & $9.2\times10^{-4}$ & 3 & 0.0663 & 0.00\,\% & 3 & \textbf{0.0052} & 0.00\,\% \\
8 & 0.2 &  1.97 & 0.0192 & 0.00\,\% & $2.1\times10^{-14}$ & 1 & 0.0391 & 10.41\,\% & 0.75 & \textbf{0.0084} & 0.37\,\% \\
8 & 1 &  9.84 & 0.0917 & 0.00\,\% & $1.1\times10^{-6}$ & 1.5 & 0.1419 & 0.00\,\% & 1.5 & \textbf{0.0528} & 0.00\,\% \\
8 & 5 & 49.19 & 0.1647 & 0.04\,\% & $1.5\times10^{-2}$ & 3 & 0.1115 & 0.57\,\% & 3 & \textbf{0.0156} & 1.00\,\% \\
\bottomrule
\end{tabular}
\end{table}

Several runs would read lower under a transient-minimum convention, and the difference is not cosmetic: at
$n = 4$, $\beta = 0.2$ the $K'$ run at $\sigma = 1$ dips to $0.0009$ and settles at $0.0014$, and at
$n = 8$, $\beta = 5$ the $K'$ run at $\sigma = 2$ dips to $0.0495$ and settles at $0.0654$. We report the
final value because stopping at a transient minimum presupposes an oracle that says when to stop,
and that oracle is exactly one more call to the exact reference, i.e.\ the currency R3 budgets. The same
caveat attaches to every ``gates to reach a target'' entry in Table~\ref{tab:threearm}, which records the
\emph{first} recorded point below the target: where the curve later rises back above it, the entry is a
cost conditional on a stopping rule the protocol does not supply. Five swept Lindbladian runs are of this
kind, three in the $K'$ column and two in $K$. The $n = 8$, $\beta = 5$, $\sigma = 2$ run is one of them,
and it is why Table~\ref{tab:threearm} credits that cell with $6.76\times10^{5}$ gates for
$\td \le 0.05$: that is the transient crossing of a run that settles at 0.0654, and it is cheaper than
reaching the same target on the $\sigma = 3$ run that settles below it at 0.0156.

\subsection{Ranking inversions}
\label{sec:inversions}

Table~\ref{tab:threearm} is the head-to-head comparison. Its informative content is the inversions rather
than the winners.

\begin{table}[tbp]
\centering\scriptsize
\setlength{\tabcolsep}{3.4pt}
\caption{The three-arm head-to-head, envelope frontier: cumulative two-qubit gates to reach a
trace-distance target, CNOT-ladder regime, at matched $(n,\beta,\text{target})$, minimized over each arm's
whole swept grid (Sec.~\ref{sec:tuning}). ``n/r'' means the arm was run and never reaches that target at
any budget in its swept range. Every cost model of every arm is run at every cell, $n = 8$
included, so no entry here is a not-run marker and every ``n/r'' is a measurement. ``Realizable''
restricts to the columns that are actual circuits, i.e.\ excludes the two \texttt{oracle} models. The native-$R_{ZZ}$ version of this table changes no entry in either winner column.
$^\dagger$: horizon-limited. At $n = 8$ the Lindbladian sweep is narrower and shorter than at $n = 4, 6$
(six widths at 300 cycles at $\beta = 0.2$, one at 1500 cycles at $\beta = 1$, two at 600 cycles at
$\beta = 5$), so its clock reaches $\tau = 18.75 / 375 / 150$ against 625--3000 at $n \le 6$. An ``n/r''
there is a statement about that horizon. Inversion counts over the fully swept $n \in \{4, 6\}$ subgrid,
and the budget-respecting counterpart of this table, are given in Sec.~\ref{sec:budgetfrontier}.}
\label{tab:threearm}
\begin{tabular}{@{}llrr rrrr rrr ll@{}}
\toprule
$n$ & $\beta$ & $\bt$ & target & SB hw & SB cl & DB oracle & DB pf & LD or.\  ($K$) & LD circ.\  ($K$) & LD circ.\  ($K'$) & winner & realizable \\
\midrule
4 & 0.2 &  0.95 & 0.3 & \textbf{48} & 512 & 148 & 196 & 540 & 732 & 468 & SB-hw & SB-hw \\
4 & 0.2 &  0.95 & 0.1 & \textbf{288} & 4608 & 292 & 292 & 5940 & 8052 & 4680 & SB-hw & SB-hw \\
4 & 0.2 &  0.95 & 0.05 & n/r & 8704 & \textbf{292} & 988 & 1.24e4 & 1.68e4 & 8424 & DB-or. & DB-pf \\
4 & 0.2 &  0.95 & 0.02 & n/r & 1.69e4 & \textbf{964} & 1.91e5 & n/r & n/r & 1.40e4 & DB-or. & LD-$K'$ \\
4 & 1 &  4.76 & 0.3 & n/r & 8704 & 1444 & 1.88e4 & 1080 & 1464 & \textbf{936} & LD-$K'$ & LD-$K'$ \\
4 & 1 &  4.76 & 0.1 & n/r & 2.10e4 & 4684 & 3.56e6 & n/r & 5.70e5 & \textbf{2340} & LD-$K'$ & LD-$K'$ \\
4 & 1 &  4.76 & 0.05 & n/r & 5.68e4 & 1.43e4 & 8.15e8 & n/r & n/r & \textbf{1.05e4} & LD-$K'$ & LD-$K'$ \\
4 & 1 &  4.76 & 0.02 & n/r & n/r & 6.99e4 & 2.09e11 & n/r & n/r & \textbf{5.80e4} & LD-$K'$ & LD-$K'$ \\
4 & 5 & 23.79 & 0.3 & n/r & 1.95e4 & 1.44e4 & 6.52e6 & 2.06e4 & 2.32e4 & \textbf{1.43e4} & LD-$K'$ & LD-$K'$ \\
4 & 5 & 23.79 & 0.1 & n/r & n/r & 3.54e5 & n/r & 2.29e5 & 2.58e5 & \textbf{3.81e4} & LD-$K'$ & LD-$K'$ \\
4 & 5 & 23.79 & 0.05 & n/r & n/r & 1.76e6 & n/r & 9.49e6 & 1.05e7 & \textbf{2.54e5} & LD-$K'$ & LD-$K'$ \\
4 & 5 & 23.79 & 0.02 & n/r & n/r & 8.82e6 & n/r & 1.48e7 & 1.49e7 & \textbf{7.73e6} & LD-$K'$ & LD-$K'$ \\
\midrule
6 & 0.2 &  1.46 & 0.3 & \textbf{72} & 768 & 330 & 410 & 1720 & 2056 & 1416 & SB-hw & SB-hw \\
6 & 0.2 &  1.46 & 0.1 & n/r & 1.15e4 & \textbf{690} & 1994 & 2.41e4 & 2.88e4 & 1.84e4 & DB-or. & DB-pf \\
6 & 0.2 &  1.46 & 0.05 & n/r & 2.07e4 & \textbf{690} & 6410 & 6.88e4 & 5.96e4 & 3.12e4 & DB-or. & DB-pf \\
6 & 0.2 &  1.46 & 0.02 & n/r & 4.22e4 & \textbf{4374} & 3.89e5 & n/r & n/r & 5.66e4 & DB-or. & SB-cl \\
6 & 1 &  7.30 & 0.3 & n/r & 2.23e4 & 4022 & 7.50e5 & 5160 & 6168 & \textbf{2832} & LD-$K'$ & LD-$K'$ \\
6 & 1 &  7.30 & 0.1 & n/r & 5.61e4 & 2.93e4 & 2.00e9 & n/r & n/r & \textbf{8496} & LD-$K'$ & LD-$K'$ \\
6 & 1 &  7.30 & 0.05 & n/r & n/r & 1.46e5 & 4.01e11 & n/r & n/r & \textbf{2.79e4} & LD-$K'$ & LD-$K'$ \\
6 & 1 &  7.30 & 0.02 & n/r & n/r & \textbf{6.98e5} & n/r & n/r & n/r & 1.04e6 & DB-or. & LD-$K'$ \\
6 & 5 & 36.48 & 0.3 & n/r & 4.45e4 & 7.23e5 & n/r & 8.90e4 & 9.91e4 & \textbf{4.19e4} & LD-$K'$ & LD-$K'$ \\
6 & 5 & 36.48 & 0.1 & n/r & n/r & 1.72e7 & n/r & 7.16e6 & 7.57e6 & \textbf{1.68e5} & LD-$K'$ & LD-$K'$ \\
6 & 5 & 36.48 & 0.05 & n/r & n/r & 8.60e7 & n/r & n/r & n/r & \textbf{4.00e5} & LD-$K'$ & LD-$K'$ \\
6 & 5 & 36.48 & 0.02 & n/r & n/r & 2.15e9 & n/r & n/r & n/r & \textbf{5.66e6} & LD-$K'$ & LD-$K'$ \\
\midrule
8 & 0.2 &  1.97 & 0.3 & \textbf{96} & 1024 & 520 & 744 & 3304$^\dagger$ & 3912$^\dagger$ & 2680$^\dagger$ & SB-hw & SB-hw \\
8 & 0.2 &  1.97 & 0.1 & n/r & 2.25e4 & \textbf{1248} & 9944 & 5.62e4$^\dagger$ & 6.65e4$^\dagger$ & 4.02e4$^\dagger$ & DB-or. & DB-pf \\
8 & 0.2 &  1.97 & 0.05 & n/r & 3.89e4 & \textbf{3848} & 5.79e5 & 1.68e6$^\dagger$ & 7.14e5$^\dagger$ & 6.43e4$^\dagger$ & DB-or. & SB-cl \\
8 & 0.2 &  1.97 & 0.02 & n/r & 1.01e5 & \textbf{7496} & 6.08e5 & n/r$^\dagger$ & n/r$^\dagger$ & 1.07e5$^\dagger$ & DB-or. & SB-cl \\
8 & 1 &  9.84 & 0.3 & n/r & 4.40e4 & \textbf{2.05e4} & 4.35e7 & 1.10e6$^\dagger$ & 1.18e6$^\dagger$ & 7.27e5$^\dagger$ & DB-or. & SB-cl \\
8 & 1 &  9.84 & 0.1 & n/r & \textbf{1.28e5} & 2.34e5 & 5.67e11 & n/r$^\dagger$ & n/r$^\dagger$ & 1.29e6$^\dagger$ & SB-cl & SB-cl \\
8 & 1 &  9.84 & 0.05 & n/r & n/r & \textbf{1.11e6} & n/r & n/r$^\dagger$ & n/r$^\dagger$ & n/r$^\dagger$ & DB-or. & none \\
8 & 1 &  9.84 & 0.02 & n/r & n/r & \textbf{5.55e6} & n/r & n/r$^\dagger$ & n/r$^\dagger$ & n/r$^\dagger$ & DB-or. & none \\
8 & 5 & 49.19 & 0.3 & n/r & \textbf{8.91e4} & 5.55e6 & n/r & 4.98e5$^\dagger$ & 5.51e5$^\dagger$ & 2.55e5$^\dagger$ & SB-cl & SB-cl \\
8 & 5 & 49.19 & 0.1 & n/r & n/r & 6.86e8 & n/r & n/r$^\dagger$ & n/r$^\dagger$ & \textbf{4.50e5}$^\dagger$ & LD-$K'$ & LD-$K'$ \\
8 & 5 & 49.19 & 0.05 & n/r & n/r & 3.43e9 & n/r & n/r$^\dagger$ & n/r$^\dagger$ & \textbf{6.76e5}$^\dagger$ & LD-$K'$ & LD-$K'$ \\
8 & 5 & 49.19 & 0.02 & n/r & n/r & 8.56e10 & n/r & n/r$^\dagger$ & n/r$^\dagger$ & \textbf{9.46e6}$^\dagger$ & LD-$K'$ & LD-$K'$ \\
\bottomrule
\end{tabular}
\end{table}

The first inversion axis is the accuracy target. Holding $(n,\beta)$ fixed and tightening the target changes the winner in six of the nine columns:
\begin{center}\small
\begin{tabular}{@{}llllll@{}}
\toprule
column & $\bt$ & $\td \le 0.3$ & $\le 0.1$ & $\le 0.05$ & $\le 0.02$ \\
\midrule
$n=4$, $\beta=0.2$ & 0.95  & SB-hw & SB-hw & DB-oracle & DB-oracle \\
$n=6$, $\beta=0.2$ & 1.46  & SB-hw & DB-oracle & DB-oracle & DB-oracle \\
$n=8$, $\beta=0.2$ & 1.97  & SB-hw & DB-oracle & DB-oracle & DB-oracle \\
$n=6$, $\beta=1$   & 7.30  & LD-$K'$ & LD-$K'$ & LD-$K'$ & DB-oracle \\
$n=8$, $\beta=1$   & 9.84  & DB-oracle & SB-cl & DB-oracle & DB-oracle \\
$n=8$, $\beta=5$   & 49.19 & SB-cl & LD-$K'$ & LD-$K'$ & LD-$K'$ \\
\bottomrule
\end{tabular}
\end{center}
Five of the six run in one direction: the arm that is cheapest at that cell wins loose targets (at
$\td \le 0.3$, $\beta = 0.2$ the hardware preset needs a \emph{single reset cycle}: 48 / 72 / 96 gates),
and a different arm wins tight ones. A benchmark that stops at 0.3 concludes that the sampler is an order of
magnitude cheaper. One that demands 0.02 concludes that only the other two work. Both are true
statements about the same three algorithms. The sixth column, $n = 8$, $\beta = 1$, is not monotone at all:
the sampler takes the middle target back from DB-TFD's oracle ($1.28\times10^{5}$ against
$2.34\times10^{5}$) and loses both ends, which is what a crossing of two differently-shaped curves looks
like when it is sampled at four points rather than two. Three of the six columns are at $n = 8$, where the
Lindbladian curve is horizon-limited ($\dagger$ in Table~\ref{tab:threearm}) and at $\beta = 1$ rests on a
single width. On the fully swept $n \in \{4, 6\}$ subgrid the count is 3 of 6.

The second inversion axis is the gate budget.
The fixed-budget cut corresponds to the constraint imposed by a finite coherence time. Two counts are
needed here. Reading the leader across $10^4, 10^5, 10^6, 10^7$ gates, the
leader changes at all in six of nine cells. It changes more than once in
three of nine, at $(n,\beta) = (6, 5)$, $(8, 1)$ and $(8, 5)$, and the number of
distinct leaders reaches three in two of nine, because at $n = 8, \beta = 1$ the leader
returns to the arm it started from. The three-or-more-leader cells are:

\begin{center}\small
\begin{tabular}{@{}llllll@{}}
\toprule
$n$ & $\beta$ & $10^4$ gates & $10^5$ gates & $10^6$ gates & $10^7$ gates \\
\midrule
6 & 5 & \textbf{DB-oracle} 0.6951 & \textbf{SB-cl} 0.1357 & \textbf{LD-$K'$} 0.0356 & \textbf{LD-$K'$} 0.0052 \\
8 & 5 & \textbf{DB-oracle} 0.8292 & \textbf{SB-cl} 0.2629 & \textbf{LD-$K'$} 0.0495 & \textbf{LD-$K'$} 0.0176 \\
\midrule
\multicolumn{6}{@{}l}{\emph{size-persistence run, $\beta=5$ only:}}\\
9 & 5 & \textbf{DB-oracle} 0.8894 & \textbf{SB-cl} 0.3646 & \textbf{LD-$K'$} 0.0722 & \textbf{LD-$K'$} 0.0722$^\dagger$ \\
\midrule
\multicolumn{6}{@{}l}{\emph{two leaders, changing more than once:}}\\
8 & 1 & \textbf{DB-oracle} 0.4520 & \textbf{SB-cl} 0.1162 & \textbf{DB-oracle} 0.0790 & \textbf{DB-oracle} 0.0092 \\
\multicolumn{6}{@{}l}{\emph{two leaders, changing once:}}\\
4 & 1 & \textbf{LD-$K'$} 0.0523 & \textbf{DB-oracle} 0.0074 & \textbf{DB-oracle} 0.0015 & \textbf{DB-oracle} 0.0000 \\
4 & 5 & \textbf{DB-oracle} 0.3386 & \textbf{LD-$K'$} 0.0706 & \textbf{LD-$K'$} 0.0385 & \textbf{LD-$K'$} 0.0060 \\
6 & 1 & \textbf{LD-$K'$} 0.0881 & \textbf{LD-$K'$} 0.0383 & \textbf{DB-oracle} 0.0088 & \textbf{DB-oracle} 0.0021 \\
\bottomrule
\end{tabular}
\end{center}
\noindent{\footnotesize $^\dagger$: horizon-limited. The Lindbladian run at $n = 9$ ends at $1.0\times10^6$
gates.}

At $n = 6, \beta = 5$ and at $n = 8, \beta = 5$: DB-TFD's oracle leads at $10^4$, the system--bath
\texttt{classical} preset at $10^5$, and the Lindbladian $K'$ variant at $10^6$ and $10^7$. All three arms
lead somewhere in one cell, twice. The $10^4$ leg of the $n = 8$ cell is the weakest: it is decided at
trace distance 0.83, 16\,\% of the way from the maximally mixed start at 0.99, by a $1.07\times$ margin over
the hardware preset, and with no Lindbladian point inside the budget, since that arm's first recorded
point at $n = 8$, $\beta = 5$ already costs $1.5\times10^4$ gates. Ranking by a single configuration is
falsified in three independent ways.

The $\beta = 5$ cell was rerun at $n = 9$ with the same presets, compiler and selected hyperparameters, as a
check on whether the three-leader sequence persists with size. It is not a tenth grid cell, and no count
in this section includes it. Under the envelope policy the leader sequence is the same as at $n = 6$ and
$n = 8$: DB-TFD's oracle at $10^4$, the system--bath \texttt{classical} preset at $10^5$, and the
Lindbladian $K'$ variant at $10^6$ and $10^7$. The $10^4$ leg has narrowed further: it is decided at trace
distance 0.89, about 10\,\% of the way from the maximally mixed start, by a $1.03\times$ margin over the
hardware preset, the margin over the runner-up having fallen through $1.14\times$, $1.13\times$,
$1.07\times$ and $1.03\times$ over $n = 4, 6, 8, 9$, again with no Lindbladian point inside the budget. Under the
budget-respecting policy the $10^4$ leader is the hardware preset itself, at 0.92 against 0.97 for both the
\texttt{classical} preset and DB-TFD's oracle, so at this size the $10^4$ leg is a near-tie among arms that
have barely left the start rather than a ranking. The two later legs narrow but hold: at $10^5$ the
\texttt{classical} preset reaches 0.36 against 0.59 for DB-TFD's oracle, a $1.61\times$ margin against
$2.07\times$ at $n = 8$, and at $10^6$ the Lindbladian arm reaches 0.072 against 0.19 for the
\texttt{classical} preset, $2.60\times$ against $3.33\times$. The Lindbladian arm was run to
$1.0\times10^6$ gates at $n = 9$, so its $10^7$ entry repeats the $10^6$ value. DB-TFD's product-formula
circuit, the Lindbladian $K$ circuit, the \texttt{hardware\_gauss} preset and $\beta \in \{0.2, 1\}$ were
not run at this size.

The third inversion axis is the cost model. Restricting the winner to the columns that are actual circuits changes the answer in 10 of the 36 cells,
always in the same direction, by demoting DB-TFD, whose oracle sits up to $3.0\times10^6\times$ below its own
second-order product-formula circuit on this grid. The convention behind that count is as follows. The two winner columns of Table~\ref{tab:threearm}
differ in 12 cells, of which 2 are cells where \emph{no} realizable arm reaches the target at all, so the
realizable winner is ``none''. Those two are excluded, because a demotion to nothing is a statement about
the grid's reach rather than about the ranking. Counting them, the figure is 12 of 36. The 10 are the cells
where a realizable arm exists and is a different arm. This is Sec.~\ref{sec:oraclegap} seen at the level of
the ranking.

The fourth axis is the compilation regime. Rule R5 was declared in advance on the expectation that the compilation regime would reorder the arms. On the
TFIM grid, at matched trace distance, it does not. Sec.~\ref{sec:heisenberg} finds one cell where it does,
on a second model, and Sec.~\ref{sec:sensitivity} two on this model at matched free energy, all by the
mechanism this paragraph identifies. The native-$R_{ZZ}$ regime re-prices a weight-$w$ Pauli rotation from
$2(w-1)$ to $w-1$, so \emph{any} all-Pauli circuit is rescaled by exactly one half whatever its weight
profile. The conjecture declared in advance, that the Lindbladian arm's weight-3/weight-4 jump operators would
break that uniformity and invert the near-tie cells, fails on both counts. With the source paper's own single-site
jumps $A_i = Y_i$ the coupling is $Y_i \otimes X_a$, weight 2, and forcing weight 3 and 4 by widening the
jump support leaves the measured \texttt{native/ladder} ratio at exactly $0.5000$ in every row, as it
must, since $(w-1)/2(w-1) = \tfrac12$ for all $w$. Both statements are pinned as tests.

What \emph{does} make the regime non-uniform across arms is a primitive that is not a Pauli rotation, and
this harness has exactly one: DB-TFD's multi-controlled phase, charged at $12L-32$ CNOTs on $L = 2n$ qubits
(Ref.~\cite{gluza2026double}'s own convention), which the native regime does not halve. Measured ratios are
$0.5000$ exactly for the system--bath arm and for the Lindbladian arm at every jump weight, against
$0.517$--$0.773$ over every swept DB-TFD \texttt{oracle} run and $0.517$--$0.930$ over every \texttt{pf}
run, the spread arising because the reflection's share of the circuit falls as the Hamiltonian evolution is
refined. Those two ranges are over per-run ratios evaluated at each run's final recorded point,
the convention used wherever this ratio appears below. Taken instead over every recorded point, the
\texttt{oracle} maximum is $0.776$ rather than $0.773$ and the \texttt{pf} maximum is unchanged, which
moves no threshold in what follows. The threshold argument that a flip needs DB-TFD to lead by less than the ratio of the two regime
factors is therefore \emph{not} available in the clean form ``$0.63/0.50 \approx 1.26$'': at the top of the
oracle range that ratio is $0.773/0.500 = 1.55$, and the grid's closest calls, $1.357\times$
($n = 6, \beta = 0.2, \td \le 0.1$) and $1.488\times$ ($n = 6, \beta = 1, \td \le 0.02$), both sit below
it. We therefore establish the conclusion by \emph{direct re-pricing} rather than by a threshold: rebuilding
all 36 cells from the \texttt{cumulative\_2q\_native} field of the same artifacts changes no entry in
either winner column, under the envelope frontier and under the budget-respecting one alike. The
determining quantity is the primitive, not the Pauli weight, which is why
\texttt{accounting.PRIMITIVES} catalogs primitives rather than weights. The threshold itself is real,
and Sec.~\ref{sec:heisenberg} crosses it: on the Heisenberg chain one cell has DB-TFD leading by
$1.30\times$ against a regime-factor ratio of $1.35$, and the winner flips.

\subsection{The budget-respecting frontier}
\label{sec:budgetfrontier}

Table~\ref{tab:threearm} is an \emph{envelope}: each tuned arm is credited with the best point in its whole
swept grid, chosen with knowledge of the exact answer, while the system--bath arm gets two untuned
configurations (Sec.~\ref{sec:tuning}). That is generous in a direction we can measure, so we measure it.

The budget-respecting frontier holds every hyperparameter that the 200-call tuning budget was
declared to select at the value the tuning selected, and lets only the cost axis vary. For the Lindbladian
arm that fixes $\sigma$, the coupling and the angle budget to the tuned $(0.75,0.25,1)$, $(1.5,0.5,1)$,
$(2,0.5,0.5)$ at $\beta = 0.2/1/5$, collapsing 28 swept configurations per cell at $n = 4$ to four (two
cost models $\times$ two channels) and leaving the cycle count free. For DB-TFD it fixes the angle budget,
the root ordering and the polynomial family to the tuned values, which restricts the arm to its
\texttt{poly} variant, since the tuning grid selected a polynomial family and never compared \texttt{poly}
against \texttt{vanilla}. The iteration count $k$ and the sub-stepping depth stay free because they are the
cost axis rather than tuning knobs. The system--bath arm is unchanged, being untuned already. No new
simulation is involved: this is a re-selection from the same artifacts.

Under that policy the three headline counts move as follows.

\begin{center}\small
\begin{tabular}{@{}lcccc@{}}
\toprule
 & \multicolumn{2}{c}{$n \in \{4,6,8\}$} & \multicolumn{2}{c}{$n \in \{4,6\}$ only} \\
\cmidrule(lr){2-3}\cmidrule(lr){4-5}
 & envelope & budget-respecting & envelope & budget-respecting \\
\midrule
winner changes with the accuracy target      & 6 of 9   & \textbf{8 of 9}   & 3 of 6  & 5 of 6 \\
winner changes with the cost model           & 10 of 36 & \textbf{16 of 36} & 6 of 24 & 13 of 24 \\
\quad (counting ``no realizable arm reaches it'') & 12 of 36 & 21 of 36 & 6 of 24 & 15 of 24 \\
leader changes with the gate budget          & 6 of 9   & 5 of 9            & 4 of 6  & 3 of 6 \\
\quad cells with three distinct leaders      & 2 of 9   & \textbf{3 of 9}   & 1 of 6  & 2 of 6 \\
winner changes with the compilation regime   & 0 of 36  & 0 of 36           & 0 of 24 & 0 of 24 \\
\bottomrule
\end{tabular}
\end{center}

\noindent The right-hand pair restricts to the sizes at which the Lindbladian arm was fully swept to a
converged horizon. Every inversion survives there, and the direction of every envelope-to-budget move is
the same. Under the envelope policy the $n = 8$ cells contribute 3 of the 6 target inversions and 4 of the
10 cost-model inversions, and those are the counts to read with the $\dagger$ caveat of
Table~\ref{tab:threearm}.

Every inversion result reported above is stronger under the budget-respecting policy, and the one
null result stays null. The direction is not an accident: the envelope's generosity is concentrated in the
two tuned arms, and removing it makes their curves less able to dominate a column uniformly, so winners
change more often rather than less. The single count that falls, the number of cells whose leader changes with the
budget, falls because at $n = 4$, $\beta = 1$ the Lindbladian arm at its tuned $\sigma$ never leads at all,
so that column stops inverting and simply belongs to DB-TFD throughout.

Three specific consequences follow. First, at $\beta = 5$ the Lindbladian arm's tuned $\sigma = 2$
is one grid step below the $\sigma = 3$ that minimizes its floor, and that single step is the largest effect
of post-hoc selection anywhere here: at $n = 4$ it moves the reachable floor from 0.0024 to 0.0385, a factor
of 16, and the fitted $\betaeff$ from 5.041 to 4.470. Second, N3 was not applied
(Sec.~\ref{sec:tuning}), and applying it would divide DB-TFD's counts by $1.29$--$1.57$. Since DB-TFD's
\texttt{oracle} column already wins every cell it could win, that can only widen the oracle-to-realizable
gap of Sec.~\ref{sec:oraclegap} and cannot reverse any inversion reported above. Third, the system--bath
arm is the only arm that is not credited with an envelope at all, so its wins, all of them at loose targets,
are the most conservative entries in either table.

\subsection{Method dependence of the oracle-to-realizable gap}
\label{sec:oraclegap}

This result is the one least dependent on the specific arms compared.

Two of the three arms admit the same split: the paper's own resource currency, priced by our compiler,
against a circuit that actually prepares the state. We apply it with the same harness, the same compiler
and the same accuracy targets and obtain

\begin{center}\footnotesize
\begin{tabular}{@{}llll@{}}
\toprule
arm & its paper's currency & a realizable circuit & ratio, matched $\td$ \\
\midrule
DB-TFD & queries to $e^{-iHt}$ and reflections & 2nd-order group commutator & $1$--$3.0\times10^6$ \\
local driving, $K$  & total Hamiltonian evolution time & Trotterized protocol & $1.11$--$1.36$ \\
local driving, $K'$ & total Hamiltonian evolution time & Trotterized protocol & $1.20$--$1.70$ \\
\bottomrule
\end{tabular}
\end{center}

The DB-TFD range follows one convention. The ratio is taken
cell by cell over the envelope frontier of Table~\ref{tab:threearm}: for each $(n,\beta,\text{target})$ at
which \emph{both} the \texttt{oracle} and the \texttt{pf} column reach the target, the entry is
$\text{pf}/\text{oracle}$, and the range is the minimum and maximum over those cells. On the TFIM grid that
is 22 of the 36 cells and the range is $1.00$ to $3.0\times10^6$: the minimum is at
$(n,\beta,\td) = (4, 0.2, 0.1)$, where both models cost 292 gates and the split has not yet opened, and the
maximum at $(4, 1, 0.02)$. On the Heisenberg subgrid the same convention gives 6 of 8 cells and a range of
$165$ to $2.1\times10^7$, the minimum at $(4, 1, 0.3)$ and the maximum again at $(4, 1, 0.02)$. Both
endpoints are properties of where the two curves are sampled rather than of the asymptotics: the ratio is
near unity only where the target is loose enough that a single group-commutator step suffices, and it grows
without visible bound as the target tightens. The budget-respecting policy moves the low end to $1.18$ and
leaves the high end unchanged, and the native-$R_{ZZ}$ regime widens the high end to $4.0\times10^6$.

The \texttt{oracle} column is the Trotter cost of the
queries with zero product-formula error: each $e^{\pm i\sqrt{s}H}$ is priced by our compiler while the
dynamics are applied exactly, so it is a lower bound on any circuit. The \texttt{pf} column is the group
commutator ($m = 2$) refined by exact sub-stepping, with every $e^{-iHt}$ compiled at second order, which is
one achievable circuit and not the cheapest: the source's own numerics use $m = 4$ and $m = 5$, and a
correct fourth-order formula exists (Sec.~\ref{sec:reimpl-db}, deviation Arm~2 D2). Because the cost
recursion is $(4N-1)^k$, the formula order sets the base of an exponential rather than a constant factor,
so the upper end of the range, and the statement that it grows without visible bound, describe this
second-order compilation. The DB-TFD ratio is therefore an upper bound on the algorithm's own gap, as the
\texttt{oracle} column is a lower bound on its cost. The share a higher-order formula removes is measured
only with that formula, which this grid does not use.

The largest discrepancy occurs in a single physical cell. At TFIM $n = 6$, $h = J$, $\beta = 1/J$,
$\td \le 0.05$: DB-TFD needs $1.46\times10^5$ two-qubit gates under its own query accounting and
$4.0\times10^{11}$ under a second-order product-formula circuit that actually prepares the state. Those
differ by a factor of up to 2.7 million, an upper bound set by that compilation, and any ranking statement
must specify which of the two it uses. The system--bath arm does not reach the target there, but that
clause is conditional: its classical fixed point is 0.0742 at the $\delta t = 0.25$ we ran, against the
source's 0.1, with zero tuning calls against the other arms' 160 and 180. Halving the step lowers the
$n = 4$, $\beta = 1$ floor by a third, which applied here would put it at the target
(Sec.~\ref{sec:tuning}). It would not take the cell, since the preset already spends $5.6\times10^4$
gates reaching 0.1 against the Lindbladian $K'$ circuit's $2.8\times10^4$ for 0.05, and a finer step only
raises its cost per cycle. Its hardware fixed point, 0.5339, is not in question.

The difference is structural, not accidental. The local-driving protocol's only non-free primitive \emph{is}
Hamiltonian evolution: its drive compiles exactly (not approximately) to three Pauli rotations of which two
are weight 1 and free, and its resets are free. The resource its paper counts and the resource a circuit
pays are therefore nearly the same resource, with the drive adding 11--36\,\% on top for the rewound
channel $K$ and 19--70\,\% for $K'$. The two differ because dropping the rewind removes Hamiltonian
evolution and not drive pulses, so the same absolute drive cost is carried by a smaller denominator.
Both ranges are over the whole swept $\sigma$ grid at $n = 4, 6, 8$, and both are dominated by their
$\sigma = 0.5$ endpoint at $n = 4$: excluding it, $K$ is $1.11$--$1.22$ and $K'$ is $1.19$--$1.35$. The
ratio falls with $n$ and with $\beta$, because both raise the Hamiltonian-evolution share of the cycle
against a drive whose pulse count grows only with the window. DB-TFD's dominant
primitive is a multi-controlled phase its paper never prices at all, and its group-commutator error is
$O((\sqrt{s}\|H\|)^3)$ while the polynomial's own step sizes are $O(1/\sqrt{V})$ and never small, so the
sub-stepping count grows like $\varepsilon^{-2}$ while the cost grows like $(4N-1)^k$.

\begin{center}\small
\begin{tabular}{@{}llll@{}}
\toprule
$n$ & $\beta$ & \multicolumn{2}{l}{local driving, circuit/oracle over the whole $\sigma$ grid} \\
    &         & $K$ (rewound) & $K'$ (no rewinding) \\
\midrule
4 & 0.2 & 1.213--1.356 & 1.336--1.696 \\
4 & 1   & 1.213--1.356 & 1.336--1.696 \\
4 & 5   & 1.127--1.182 & 1.224--1.348 \\
6 & 0.2 & 1.189--1.195 & 1.302--1.311 \\
6 & 1   & 1.189--1.195 & 1.302--1.311 \\
6 & 5   & 1.112--1.114 & 1.201--1.204 \\
8 & 0.2 & 1.179--1.184 & 1.287--1.293 \\
8 & 1   & 1.180        & 1.288 \\
8 & 5   & 1.106--1.107 & 1.191--1.192 \\
\bottomrule
\end{tabular}
\end{center}

The consequence for the wider literature follows directly. A literature that reports query counts is not
making a uniformly-sized error: it is making an error whose size is anticorrelated with how analog-native
the algorithm is. Block-encoding and double-bracket methods are strongly favored by their own currency, while a
repeated-interaction protocol is barely affected. Any ranking assembled from published resource statements
therefore carries a systematic bias with a known sign, and Sec.~\ref{sec:inversions} shows it changing the
answer in 10 of 36 cells.

Two measured conditions qualify this. First, an oracle model is a cost lower bound, not an
accuracy upper bound: at $n = 4$, $\beta = 1$, $\sigma = 2$ the \texttt{oracle} floor is 0.155 and the
\texttt{circuit} floor is 0.058, i.e.\ the Trotterized protocol is \emph{more} accurate there, because
Trotter error is a perturbation that can move a fixed point either way. Second, the paper's own units
travel alongside in every artifact: at $n = 4$, $\beta = 1$, $\sigma = 1.5$, $K'$ spends 2386 units of
Hamiltonian evolution time to reach $\td \le 0.018$, at cycle 270 of 12\,000, and never returns above that
target over the remaining 106\,036 units.

\subsection{Scaling in the normalized inverse temperature}
\label{sec:clock}

Sec.~\ref{sec:reference} fixed $\bt = \beta\lambda$ as the axis. Three consequences follow.

\begin{figure}[t]
\centering
\includegraphics[width=0.92\textwidth]{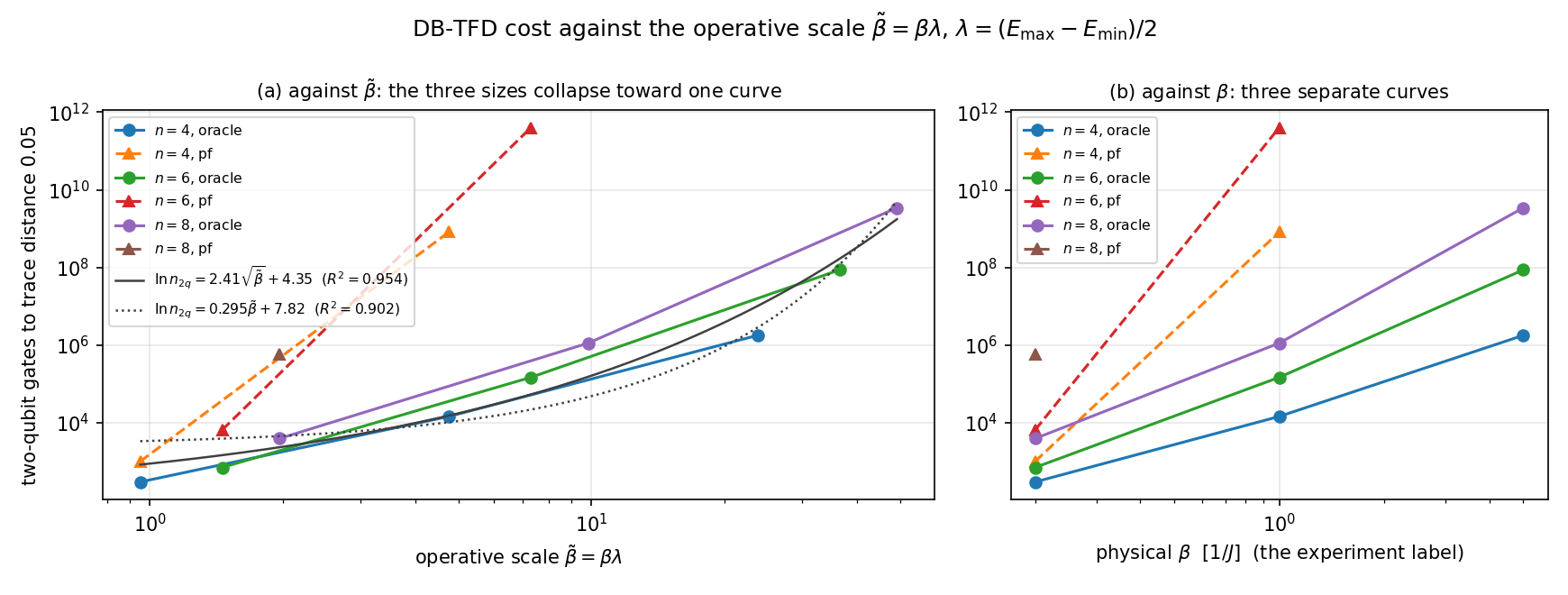}
\caption{DB-TFD cost against normalized inverse temperature. (a) Two-qubit gates to $\td \le 0.05$ against
$\bt = \beta\lambda$, the primary panel, with the two global fits over the nine oracle-model cells.
(b) The same nine numbers against physical $\beta$, the experiment label. The three sizes are
three separate curves in (b) and collapse toward one in (a). Because $\lambda$ grows linearly in $n$, what
reads as a temperature scaling in (b) is partly a size scaling. Both fits, and the leave-one-out and
bootstrap statistics quoted in Sec.~\ref{sec:sensitivity}, are computed from the same artifacts by
\texttt{beta\_scaling\_fit} in \texttt{scripts/run\_dbtfd.py}.}
\label{fig:clock}
\end{figure}

First, at fixed physical $\beta = 1/J$, going from $n = 4$ to $n = 8$ doubles $\bt$ and multiplies DB-TFD's
cost of $\td \le 0.05$ by $78\times$ ($1.43\times10^4 \to 1.11\times10^6$). DB-TFD's advertised
temperature scaling is therefore also a size scaling (Fig.~\ref{fig:clock}).

Second, the polynomial variant's iteration count is consistent with growth like $\sqrt{\bt}$: across
$\bt \in [0.95, 49.2]$, a factor of 51.7, the smallest $k$ reaching $\td \le 0.05$ is
$k = (1.03\text{--}1.66)\sqrt{\bt}$, constant to $\pm 24\,\%$ (and to $\pm 13\,\%$ once the
quantization-dominated $k \le 2$ points are dropped, Sec.~\ref{sec:sensitivity}). Fitting the cost gives
$\ln \twoq = 2.41\sqrt{\bt} + 4.35$ ($R^2 = 0.954$) against $\ln \twoq = 0.295\bt + 7.82$
($R^2 = 0.902$): $\exp(O(\sqrt{\bt}))$ in the range a benchmark can reach, $\exp(\tilde O(\beta))$
asymptotically as its Theorem 1 says. Both are true and about different regimes, a distinction that
only appears when gates are plotted instead of queries. The discrimination between the two fits is
one-signed on every subset of this grid but not decisive at 95\,\%, and Sec.~\ref{sec:sensitivity} gives
the leave-one-out and bootstrap figures rather than resting on $\Delta R^2$. The measured prefactor is
not compared to the constant in Lemma 2 of Ref.~\cite{wright2026double}: that lemma bounds
$k \ge e\sqrt{(\beta/2)\log(\cdot)}$, so its effective prefactor on $\sqrt{\beta\log(\cdot)}$ is
$e/\sqrt2 \approx 1.92$ and it carries a logarithmic factor ours does not
(Sec.~\ref{sec:reimpl-db}).

Third, for the Lindbladian arm the analogous scales are $t^\star$ and the protocol time $\tau = kJ^2$. At
$n \le 6$ every run reaches $\tau = 750$ ($\beta = 0.2$) or 3000 ($\beta = 1, 5$) at $n = 4$ and 750
or 625 at $n = 6$, which is between 3.7 and about 2200 times its own $t^\star$ wherever that is finite.
At $n = 8$ the horizon is $\tau = 18.75 / 375 / 150$ at $\beta = 0.2/1/5$ and $t^\star$ is not computed,
so no mixing-time statement is made there. A benchmark that plots against physical
$\beta$ alone mis-attributes both effects, which is why $\bt$ is the axis of every table here and the
primary axis of Fig.~\ref{fig:clock}.

\subsection{Cross-check on a second model family}
\label{sec:heisenberg}

Everything above is measured on one model, and the bias-floor taxonomy is the part of it most likely to be
a property of that model rather than of the three protocols. The isotropic Heisenberg chain
$H = J\sum_i (X_iX_{i+1} + Y_iY_{i+1} + Z_iZ_{i+1})$, open, $J = 1$, is the natural control: the harness
already supports it, it is not free-fermion, and its spectral scale is close enough to the TFIM's that the
comparison is at matched $\bt$ and not merely at matched $\beta$
($\lambda = 4.732 / 7.487$ at $n = 4/6$, against $4.759 / 7.296$, so $\bt = 4.73 / 7.49$ against
$4.76 / 7.30$). All three arms were run at $n = 4, 6$, $\beta = 1$, under the standard protocol: the same
tuning budget re-spent at $n = 4$ on this model, the same swept grids, both cost models per arm, both
compilation regimes.

\begin{table}[t]
\centering\small
\caption{The Heisenberg cross-check, $\beta = 1$, alongside the TFIM cells at the same $(n,\beta)$.
Floors are final recorded trace distances, the convention of Table~\ref{tab:floors}. Every Heisenberg
cell has last-decile drift of at most 0.01\,\%. The system--bath and DB-TFD geometries carry over unchanged.
The Lindbladian arm acquires a fourth floor mechanism that the TFIM grid cannot exhibit.}
\label{tab:heisenberg}
\setlength{\tabcolsep}{4.5pt}
\begin{tabular}{@{}llrrrrrrr@{}}
\toprule
& & & \multicolumn{2}{c}{SB \texttt{hardware}} & \multicolumn{2}{c}{SB \texttt{classical}}
& DB-TFD & LD $K'$ \\
\cmidrule(lr){4-5}\cmidrule(lr){6-7}\cmidrule(lr){8-8}\cmidrule(lr){9-9}
model & $n$ & $\bt$ & $\td$ & $\betaeff$ & $\td$ & $\betaeff$ & oracle, $k=12$ & floor \\
\midrule
Heisenberg & 4 & 4.73 & 0.5394 & 0.253 & 0.0410 & 0.894 & $2.9\times10^{-10}$ & 0.5009 \\
Heisenberg & 6 & 7.49 & 0.6517 & 0.237 & 0.0327 & 0.928 & $7.3\times10^{-8}$  & 0.4724 \\
\midrule
TFIM       & 4 & 4.76 & 0.4467 & 0.325 & 0.0494 & 0.902 & $2.1\times10^{-10}$ & 0.0176 \\
TFIM       & 6 & 7.30 & 0.5339 & 0.317 & 0.0742 & 0.888 & $4.2\times10^{-8}$  & 0.0158 \\
\bottomrule
\end{tabular}
\end{table}

The system--bath and double-bracket geometries carry over. Table~\ref{tab:heisenberg} sets the
Heisenberg cells against the TFIM cells at the same $(n,\beta)$. The system--bath arm's temperature ceiling
is there and is lower: at the 139-qubit preset $\betaeff = 0.253 / 0.237$ at $n = 4/6$ for requested
$\beta = 1$, against $0.325 / 0.317$ on TFIM, at zero drift, while the same code at the classical preset
calibrates correctly ($\betaeff = 0.894 / 0.928$, relative free-energy error
$1.9\times10^{-3}$ / $1.1\times10^{-3}$). DB-TFD again has no floor, only a budget: its \texttt{oracle}
model reaches $2.9\times10^{-10}$ and $7.3\times10^{-8}$ at the grid ceiling $k = 12$, while its
realizable product formula stalls at $1.7\times10^{-2}$ and $5.2\times10^{-2}$. The oracle-to-realizable
gap is the same phenomenon at the same magnitude and with the same sign: DB-TFD needs
$1.11\times10^{4}$ two-qubit gates for $\td \le 0.05$ at $n = 4$ under its own query accounting and
$9.17\times10^{8}$ as a circuit, a factor of $8.3\times10^{4}$, while the local-driving arm's
circuit-to-oracle ratio is $1.06$--$1.07$ ($K$) and $1.11$--$1.12$ ($K'$), i.e.\ closer to one than on
TFIM.

The Lindbladian arm shows a fourth floor mechanism. On Heisenberg the arm floors at $0.5009$ at $n = 4$ and $0.4724$ at $n = 6$, and unlike on TFIM the floor
barely moves with $\sigma$ below $\sigma^\star$: $0.5009 / 0.5009 / 0.5018$ at $\sigma = 0.5 / 0.75 / 1$ at
$n = 4$, and $0.4739 / 0.4724 / 0.4724$ at $n = 6$. The cause is the jump family. App.~E of Ref.~\cite{hahn2025towards} takes $A_i = Y_i$, one Pauli letter
translated along the chain, so $S_Y = \sum_i Y_i$ commutes with every $A_a$ \emph{by construction}. On a
Hamiltonian that also commutes with $S_Y$, which the isotropic chain does exactly, the filtered jump
operators $L_a$ and the coherent term $G$ inherit that commutation, and the exact
KMS-detailed-balanced sampler stops being ergodic: at $n = 4$ its generator has a five-dimensional kernel,
one stationary state per $S_Y$ eigenvalue, against a one-dimensional kernel on TFIM at $h = J$ with a
spectral gap of $6.3\times10^{-2}$. The Gibbs state remains a fixed point but is no longer the only one,
and uniqueness is what makes the protocol a sampler.

The consequence is predictable in closed form and matches. A protocol that cannot move population between
$S_Y$ sectors, started from the maximally mixed state, can at best reach
$\rho_\star = \sum_s (p_s^{\mathbb 1}/p_s^{\beta})\,\Pi_s\rho_\beta\Pi_s$ with $\Pi_s$ the projector onto
the $s$-th sector, and $\tfrac12\|\rho_\star - \rho_\beta\|_1$ is $0.5009$ at $n = 4$ and $0.4724$ at $n = 6$,
which is the measured floor \emph{to four decimals at both sizes}. The obstruction is pinned as a test
(\texttt{test\_single\_letter\_jumps\_are\_not\_ergodic\_on\_an\_su2\_sym\-metric\_chain}). It is a
property of the jump set rather than of the protocol, and it is invisible on TFIM at $h = J$ and on
Ref.~\cite{hahn2025towards}'s own mixed-field Ising model, where no such conservation law exists. It is
the fourth distinct floor mechanism reported here and the only one a single model family could not have
produced.

On the second model the cost-model inversion
strengthens: restricting the winner to realizable circuits changes the answer in 7 of the 8
Heisenberg cells against 10 of 36 on TFIM, always by demoting DB-TFD, because on this model the
realizable arms are weaker and the oracle column is further ahead. The accuracy-target inversion does not
appear in these two columns (0 of 2), for a reason the table makes plain: with the Lindbladian arm held at
a symmetry floor above every target, DB-TFD's \texttt{oracle} column wins all eight cells outright and
there is no second arm to cross it.

The compilation-regime result changes, in the direction Sec.~\ref{sec:inversions} predicted and could not
demonstrate. On TFIM the regime reorders nothing in 36 cells. On Heisenberg it reorders one of
eight, at $n = 6$, $\td \le 0.1$: under the CNOT ladder DB-TFD's \texttt{oracle} needs
$1.106\times10^{5}$ gates against the system--bath \texttt{classical} preset's $1.434\times10^{5}$ and
leads by $1.30\times$. Under native $R_{ZZ}$ the two become $7.46\times10^{4}$ and $7.17\times10^{4}$ and
the sampler leads by $1.04\times$. The mechanism is exactly the one Sec.~\ref{sec:inversions} identifies:
the system--bath count halves exactly, DB-TFD's does not, because its multi-controlled phase is not a
Pauli rotation. Its measured \texttt{native/ladder} ratio at that cell is $0.675$, so the flip needed a
lead below $0.675/0.500 = 1.35$, and $1.30$ is below it. Rule R5's declared axis does move a ranking, and
a second model family is required to exhibit it. The threshold argument the TFIM grid could only state is
here realized.

The floor taxonomy survives the change of model for two arms of three and is
\emph{extended} by the third: bias floor, no floor, tunable floor, and now ergodicity floor. The
oracle-to-realizable gap, the result Sec.~\ref{sec:discussion} claims should transfer, transfers with its
sign, its method dependence and its order of magnitude intact. The ranking-inversion counts are
properties of the grid and do not transfer as numbers: two of the three inversion axes move in opposite
directions between the two models, which is itself the reason the paper reports a frontier and not a
winner.

\subsection{Declared side channels}
\label{sec:sidechannels}

Rule R1 exists because a two-qubit-gate count hides everything in this table, and the three arms differ in
every column.

\begin{center}\footnotesize
\begin{tabular}{@{}lrrrrr@{}}
\toprule
arm ($n = 4/6/8$) & qubits & ancillas & resets / unit & Pauli strings measured / unit & native/ladder \\
\midrule
system--bath & 5 / 7 / 9 & 1 & 1 per cycle & 0 & 0.5000 \\
DB-TFD \texttt{oracle} & 8 / 12 / 16 & 0 & 0 & 28 / 66 / 120 (\texttt{poly}) & 0.517--0.773 \\
DB-TFD \texttt{pf} & 8 / 12 / 16 & 0 & 0 & 28 / 66 / 120 (\texttt{poly}) & 0.517--0.930 \\
local driving & 8 / 12 / 16 & 4 / 6 / 8 & 4 / 6 / 8 per cycle & 0 & 0.5000 \\
\bottomrule
\end{tabular}
\end{center}

DB-TFD's ``no ancillas, no post-selection'' claim survives compilation, but is paid for with twice the
qubits, and its \texttt{poly} variant additionally estimates $|H| + |H^2|$ Pauli strings every iteration
(its \texttt{vanilla} variant estimates none, and is otherwise identical in this table).
The Lindbladian arm is the most reset-hungry of the three by a factor of $n$ and the only one needing
\emph{zero} Hamiltonian-term measurements. On a device where qubits are scarce the comparison at $n = 8$
is 16 against 9, and where mid-circuit reset is the bottleneck it is 8 per cycle against 1. None of that is in
a gate count.

\subsection{Determinism, sensitivity, and the limits of the fits}
\label{sec:sensitivity}

Determinism comes first. Every run reported here is at
\texttt{seed = 0}, and the double-bracket and Lindbladian artifacts carry no seed field at all. That is
defensible only because none of the three arms is stochastic \emph{as simulated}. The system--bath arm's
randomization over $(\text{site}, \text{Pauli}, \omega)$ is state-independent, so its shot-averaged density
matrix evolves under the exactly-averaged channel of Eq.~(2) of Ref.~\cite{ding2025simple}, evaluated over
all $3n$ (site, Pauli) choices and a fixed quadrature of $g(\omega)$ (deviation Arm~1 D2). The Lindbladian
arm traces out its ancillas exactly, and DB-TFD is a deterministic sequence of unitaries and reflections,
with shot noise on its energy and variance estimates not simulated (deviation Arm~2 D7). The only place a
seed can enter is the \texttt{omega\_mode = "random"} option of the system--bath arm, which the grid does
not use. Repeating any cell at another seed reproduces it bit for bit, so error bars over seeds
would be identically zero and are not reported. The uncertainty that does exist is not sampling
uncertainty: it is the discreteness of the swept grids and the choice of selection policy, and those are
what Sec.~\ref{sec:budgetfrontier} varies instead.

The two fits of Sec.~\ref{sec:clock}, in $\sqrt{\bt}$ and in $\bt$, are over nine points and differ by
$\Delta R^2 = 0.053$ ($R^2 = 0.954$ against $0.902$). With nine points $R^2$ alone does not decide, so we
make three further checks.
Leave-one-out prefers $\sqrt{\bt}$ in 9 of 9 subsets, with $\Delta R^2$ between $+0.023$ and
$+0.099$. The leave-one-out predictive error is twice as large for the linear fit
($\mathrm{PRESS} = 33.2$ against $16.6$, i.e.\ $\mathrm{RMSE}_{\mathrm{loo}} = 1.92$ against $1.36$ in
$\ln \twoq$). A 10\,000-resample \emph{pairs} bootstrap, resampling the nine $(\bt, \ln\twoq)$ pairs
with replacement at a fixed seed and discarding any resample with fewer than three distinct pairs (on which
both fits are exact and $\Delta R^2$ undefined), puts the $\sqrt{\bt}$ slope at $2.41$ with a 95\,\%
percentile interval $[2.00, 3.29]$ and the median $\Delta R^2$ at $+0.048$ with interval
$[-0.017, +0.170]$, i.e.\ $P(\Delta R^2 > 0) = 0.91$. The evidence is consistent and one-signed but not
decisive at 95\,\%, and the supportable statement is that $\sqrt{\bt}$ fits better on every subset of this
grid, not that $\bt$ is excluded. Nine $(n,\beta)$ points spanning a factor of 51.7 in $\bt$ is the most
this grid can offer. Separating $\exp(O(\sqrt{\bt}))$ from $\exp(\tilde O(\bt))$ decisively needs a wider
$\bt$ range than an exact simulator reaches.

Grid granularity limits one statement. The constancy of
$k/\sqrt{\bt} = 1.03$--$1.66$ (Sec.~\ref{sec:clock}) is measured on an integer-valued $k$ over a grid whose
smallest point is $\bt = 0.95$, where $k = 1$. At that end a single unit of $k$ is the whole quantity, so
part of the $\pm 24\,\%$ spread is quantization rather than physics: dropping the three points with
$k \le 2$ leaves $1.23$--$1.59$, a $\pm 13\,\%$ spread over the remaining six. Both numbers are given
because neither alone is representative.

Rule R4 also allows the head-to-head on free energy. Every artifact stores the relative free-energy error
beside the trace distance, so the comparison can be re-selected on it without simulation. At the decade
targets $\Delta F/F \le 10^{-1}, 10^{-2}, 10^{-3}, 10^{-4}$, chosen for this check rather than in advance,
the counts become, envelope then budget-respecting: winner changes with the target in 7 and 7 of 9
columns (6 and 8 on trace distance), with the cost model in 15 and 17 of 36 cells (10 and 16) and with the
compilation regime in 2 and 2 of 36 (0 and 0). The leader changes with the budget in 6 and 5 of 9
(unchanged) and has three distinct leaders in 1 and 3 of 9 (2 and 3). The two regime flips are near-ties at the
$10^{-2}$ target in which DB-TFD's \texttt{oracle} narrowly leads an all-Pauli arm: under the envelope,
292 against the hardware preset's 336 at $(n,\beta) = (4, 0.2)$, which native re-pricing turns into 184
against 168, and 1636 against $K'$'s 1872 at $(4, 1)$, which becomes 964 against 936. The budget-respecting
pair is $(4, 0.2)$ and $(6, 0.2)$ against the hardware preset. That is the threshold mechanism of
Sec.~\ref{sec:inversions}, which the free-energy targets reach on this model. Every inversion axis
therefore survives the change of metric, and the TFIM zero for the compilation regime is a property of
matching in trace distance as well as of the grid. The re-selection is
\texttt{paper/free\_energy\_reselect.py}, which reuses the selection code of
\texttt{scripts/make\_paper\_tables.py} unchanged.

\section{Discrepancies surfaced by reimplementation}
\label{sec:reimpl}

We reimplemented all three protocols from their published descriptions, in two cases without
reference code and in the third without reliance on an external transpiler. Each source contained at
least one issue that materially affects a reimplementation and is not apparent from the published text.
Each is stated below with the regime in which it matters and its measured consequence, and each is
pinned as a regression test. The two items we call errors are typographical, and we found no result
either source reports that depends on them. In two of the three cases the published numerics also lie
outside the asymptotic regime the corresponding theorems address, the usual gap between a theorem and
its numerics rather than a defect of either paper, which one of the two sources states itself. Each arm additionally carries a deviation list, every entry of
which is stated with its measured effect and its sign (App.~\ref{app:deviations}: 7, 9 and 10 deviations
for the three arms in turn).

\subsection{The system--bath protocol}
\label{sec:reimpl-sb}

Four findings follow, none stated in Ref.~\cite{ding2025simple} or Ref.~\cite{farrell2026preparing}.
The versions cited are recorded in the reference list.

The first finding is a Nyquist cliff rather than a Trotter drift. Trotterizing the bath Hamiltonian
$H_E = -\omega Z_E/2$ samples the bath phase at rate $1/\delta t$, so bath frequencies above $\pi/\delta t$
alias onto lower ones and detailed balance is destroyed. With $\omega_{\max} = 8$ this requires
$\delta t < \pi/8 = 0.393$, and crossing it is a cliff: the steady-state trace distance is 0.0494 at
$\delta t = 0.25$, 0.4688 at 0.5, and 0.7089 at $\delta t = 1$, exactly the distance the
maximally mixed state already had, i.e.\ the protocol has ceased to act at all. Per-cycle
residuals of $10^{-14}$--$10^{-9}$ show these are genuine fixed points, not slow convergence, and sliding
$\omega_{\max}$ across the bound at two step sizes confirms aliasing rather than ordinary Trotter error:
at $\delta t = 0.25$ (bound 12.57) the error is flat in $\omega_{\max}$ to within 7.4\,\% across a factor
of three, while at $\delta t = 0.5$ (bound 6.28) the same sweep degrades $5.1\times$ and keeps degrading.
The condition is stated nowhere in either paper, confirmed by exhaustive search of both full
sources for any sampling, aliasing or Nyquist condition relating $\delta t$ to $\omega_{\max}$. The
closest either comes is an ordinary Trotter-error bound that happens to contain $\omega_{\max}$ as a norm.
The experiment's $(0.25, 8)$ satisfies the bound with 57\,\% margin, so the condition was met there, but
that margin is thinner than the strict inequality suggests ($\omega_{\max} = 6$ at $\delta t = 0.5$ is
nominally safe and already $4.5\times$ worse than the same $\omega_{\max}$ at $\delta t = 0.25$). Any
benchmark that coarsens $\delta t$ for a ``generous compilation'' regime must shrink $\omega_{\max}$ or
it silently changes the fixed point. \texttt{SystemBathChannel} emits a warning in this regime.

The second finding attributes the hardware preset's failure to the window rather than the coupling. The hardware preset's failure would naturally
be attributed to its strong coupling $\alpha = 1.75$. It is not: at $T = 8$ that same $\alpha$ gives
trace distance 0.0946, $1.9\times$ worse than $\alpha = 1$'s 0.0494 and nowhere near the 0.4467 the
hardware preset produces at $n = 4$, $\beta = 1$. Nor is it the envelope \emph{shape}: replacing the
constant envelope by a Gaussian of width $\sigma T = 0.19$ moves that number to 0.4348, a 2.7\,\% change,
while the classical preset's window is 10.7 times longer. What the short window does is destroy the
filter's spectral resolution, which for a window of length $T$ is $O(1/T)$ whatever its shape, so the
bath cannot resolve one Bohr frequency of the model from another and the detailed-balance factor
$e^{-\beta\omega}$ is smeared across the window. The controlled envelope sweep, at fixed
$\sigma = 0.25$ and varying $T$, gives 0.1027 / 0.0494 / 0.0424 / 0.0453 / 0.0703 at
$\sigma T = 1/2/3/4/6$ (the rise at 6 is window truncation, not physics). Neither paper frames $T$ this
way: Ref.~\cite{ding2025simple} states only $T = \tilde\Omega(\sigma)$, and
Ref.~\cite{farrell2026preparing} chooses $T = 0.75$ ``to balance quantum resources and fixed point
error''.

The third finding is the growth of the mixing time with $n$ at a single ancilla. Cycles for the \texttt{classical} preset's
trace distance to come within 1\,\% of its own final value, measured at all nine cells, the
$n = 8$ runs being converged: 73 / 96 / 122 for $n = 4/6/8$ at $\beta = 0.2$, 113 / 143 / 181 at
$\beta = 1$ and 152 / 191 / 227 at $\beta = 5$. The trajectories are monotone, so the first-crossing and
the stay-within readings of that convention give the same cycle at all nine cells. One reset cycle kicks
one randomly chosen site, so the cycle count grows roughly linearly in $n$ at every $\beta$: the
$n = 4 \to 8$ increase is $1.67\times$ at $\beta = 0.2$, $1.60\times$ at $\beta = 1$ and $1.49\times$ at
$\beta = 5$, all below the factor of two in system size. The slope does not grow with $\beta$. It
falls slightly across the three temperatures, which do not fix its direction. The 139-qubit experiment reports size-\emph{independent} depth to
equilibrium, but it runs
$N_E = 60$ environment qubits against 79 system spins, so up to 60 of the 79 sites, 76\,\%, are kicked
every cycle rather than one.
Size independence is therefore a property of the ancilla budget rather than of the algorithm, which is
why R1 exists.

The fourth finding is the non-overlap of the theoretical and experimental regimes. Ref.~\cite{ding2025simple} proves its fixed point for
$\alpha \ll 1$, $\alpha T \ll 1$, and its Theorem~S4 makes the requirement explicit: it sets $\alpha = \tilde\Theta(\omega_{\max}\beta^{-1}t_{\mathrm{mix}}^{-3/2}\epsilon^{3/2})$
alongside $\sigma = \tilde\Theta(\omega_{\max}^{-1}\beta\epsilon^{-1}t_{\mathrm{mix}})$, so $\alpha$ must
\emph{shrink} as the target error does. With the $\int f^2 dt = 1$ normalization the per-cycle transition
probability is $O(\alpha^2)$, so the experiment's $\alpha = 1.75$ means $\alpha^2 \approx 3$ and even the
classical benchmark uses $\alpha = 1$. Everything numerical in this arm, the source experiment included,
sits outside the asymptotic regime the theorem addresses, which the experiment does not claim to occupy.
The measured
price is a factor $\approx 1.9$ in trace distance, which is real but secondary to the window effect of the second finding.

The four findings are reproduced across six configurations by a
one-parameter-at-a-time sweep from the \texttt{classical} base configuration at 300 cycles, seed 0. It was
run at TFIM $n = 4$ for $\beta = 0.2, 1, 5$, at TFIM $n = 6$ for $\beta = 0.2, 1$, and on the Heisenberg
chain at $n = 4$, $\beta = 1$: two sizes, three temperatures, two models. Every effect is present at every
point with the same sign. One of the four moves in size.

At every point, crossing
  $\delta t = \pi/\omega_{\max} = 0.393$ multiplies the steady-state trace distance by $6.4$--$10.9\times$
  at the five TFIM points and by $17.2\times$ on Heisenberg (0.0407 at $\delta t = 0.25$ against 0.7006 at
  0.5), while halving $\delta t$ \emph{below} the bound changes it by at most $1.62\times$ anywhere, the
  largest being $0.0164 \to 0.0101$ at $n = 6$, $\beta = 0.2$. The
  location of the cliff does not move at all, which is what an arithmetic bound predicts and what a
  Trotter-error argument would not.

In the aliasing control, sliding $\omega_{\max}$ over a factor of
  three at $\delta t = 0.25$ (bound 12.57) moves the fixed point by 7--35\,\% at the TFIM points and
  29\,\% on Heisenberg. The same sweep at $\delta t = 0.5$ (bound 6.28) degrades it by
  $4.2$--$7.3\times$ and $5.7\times$. Ordinary Trotter error would grow at both step sizes.

For the window-versus-coupling attribution, raising $\alpha$ from 1 to 1.75
  at $T = 8$ costs $1.7$--$2.1\times$ at the TFIM points and $4.0\times$ on Heisenberg, against the
  $9.0\times$ the hardware preset's short window costs at TFIM $n = 4$, $\beta = 1$. The margin is
  narrowest on the model where the coupling matters most, and the attribution survives it there.

The window optimum is the one quantity that moves with the configuration: $\sigma T^\star$ is 4 at $\beta = 0.2$
  and 3 at $\beta = 1$, at both TFIM sizes, 3 at $\beta = 5$, which was studied at $n = 4$ only, and 2 on
  Heisenberg: it moves with the temperature and the model and, over the two sizes where both are available,
  not with the size. The \emph{mechanism} (spectral resolution $O(1/T)$) transfers. The optimal
  $T$ does not.

The two source papers disagree with each other on the window, in an ambiguity that can silently double
every gate count in this arm: Ref.~\cite{ding2025simple} writes the evolution over $[-T,T]$ with
$M = \lceil 2T/\tau\rceil$ steps, a window of length $2T$, while Ref.~\cite{farrell2026preparing} writes
$T/\delta t$ steps at midpoints $-T/2 + (j+\tfrac12)\delta t$, a window of \emph{total} length $T$. The same
symbol therefore means two different durations in the theory paper and its own companion experiment. We
follow the experiment, the source of the numbers.

No hyperparameter of this arm is inferred. Methods~D of Ref.~\cite{farrell2026preparing} states the
hardware filter width of the 139-qubit run outright as $\sigma = \infty$, a constant envelope, and its
Methods~A repeats it. The $\sigma^2 = 1/16$ that appears in that paper belongs to
a noiseless $N_S = 12$, $N_E = 1$ statevector benchmark at $T = 8$, $\delta t = 0.1$, $\alpha = 1$, which is
its \texttt{classical} setting and not its hardware run. Every \texttt{hardware} number in this work is
therefore produced at $\sigma = \infty$. The Gaussian-envelope control runs are retained in the artifact under the
label \texttt{hardware\_gauss} as the sensitivity point behind the window-versus-coupling finding: at $T = 0.75$ the fixed point moves
very little with the envelope \emph{shape}, which is why the item blames the window and not the width.

\subsection{The double-bracket protocol}
\label{sec:reimpl-db}

Throughout this subsection $\bt$ is
\emph{our} notation: Ref.~\cite{wright2026double} normalizes $\|H\|_{\mathrm{op}} \le 1$, so its $\beta$ is already
the normalized inverse temperature and it never writes a $\bt$.

Eq.~(G9) contains a transcription error that is absent from the source it cites.
  App.~G presents a six-factor product formula claimed to be $O(s^4)$. Transcribed verbatim and measured on
  random anti-Hermitian generators it is $O(s^3)$ (fitted exponent $2.99$), the same order as the
  plain group commutator of its Eq.~(8) and with an error constant $3.24\times$ larger, so as printed it is
  strictly worse than the four-factor formula it is meant to improve on. The defect traces to the
  source the paper cites for it. Eq.~(8) of Ref.~\cite{chen2022efficient} is the same six-factor template
  with the same sign pattern, but its coefficients are built from the golden-ratio \emph{conjugate}
  $(\sqrt5-1)/2 = 0.618\ldots$, where (G9) writes them in terms of $\phi = (1+\sqrt5)/2 = 1.618\ldots$.
  Substituting $\phi \to \phi - 1$ into the printed template reproduces the source coefficients exactly, and
  the source formula measures order $4.00$. The defect is therefore a transcription slip in
  Ref.~\cite{wright2026double} rather than an error inherited from Ref.~\cite{chen2022efficient}, and it is
  repairable in one character. Both statements are pinned as tests
  (\texttt{PRODUCT\_FORMULA\_ORDER\_VERIFIED}, \texttt{CHEN\_THIRD\_ORDER\_FACTORS}).

  Whether the paper's $m = 4/5$ numerics inherit the slip depends on which formula seeds the recursions
  (G10)/(G11), which the paper does not state: Ref.~\cite{wright2026double} states only that its Fig.~2
  uses ``a fourth-order product formula'' and that Table~I and its generative-modeling runs use $m = 4$
  and $m = 5$. Those recursions have no stated base case, and (G8)/(G9) are the only formulas the paper
  supplies. \emph{If} the $m = 4/5$ numerics were seeded with (G9) as printed they would inherit the slip,
  and if the intended constant was the source's they are unaffected. Our own choice does not depend on
  the resolution: we use only
  the group commutator ($m = 2$) and refine by exact sub-stepping, which needs no higher-order formula at
  all, because with $\Psi$ frozen within a step $(e^{(s/N)[\Psi,H]})^N = e^{s[\Psi,H]}$ exactly.

App.~G2 contradicts main-text Eq.~(8). Eq.~(8) writes the group commutator with
  $e^{i\sqrt{s}H}$ leftmost, i.e.\ $A = iH$, $B = i\Psi$. App.~G2 says $A$ and $B$ are replaced by
  $i\psi$ and $iH$ respectively, i.e.\ the opposite, and its Eq.~(G12) confirms that ordering. Under
  $A = i\psi$, $B = iH$ the group commutator generates $e^{-s^2[\Psi,H]}$, the opposite sign, so the two
  readings are not interchangeable without a sign flip on $s$, which matters because the polynomial variant's
  step sizes are negative by construction. We follow the main text and absorb a negative $s$ by flipping the
  sign of the Hamiltonian evolution, which also keeps the trailing $\Psi$-factor a global phase.
  That last point determines the cost of the appendix reading. The paper's own recursion base is $3^k$
  precisely because one of the four factors is free: with $|\psi_i\rangle$ the current state,
  $e^{ia\Psi_i}|\psi_i\rangle = e^{ia}|\psi_i\rangle$, so a $\Psi$-factor at the \emph{right} end of the
  product costs nothing, and the step invokes $V_i$ three times rather than five. Order the factors as
  App.~G2 does and the trailing factor is a Hamiltonian evolution instead, no reflection is free, the
  recursion base becomes 5, and the depth is multiplied by $(5/3)^k$: $7.7\times$ at $k = 4$ and
  $59.5\times$ at $k = 8$. The paper states the base 3; the base 5 follows from its construction.

The paper's resource table has no gate counts. Its Table~II compares eight
  algorithms by \emph{asymptotic} query complexity. Its ``Classical costs'' column is populated for exactly
  one of the eight rows, the classically assisted QITE entry, at $O(\exp(C(\beta)^d))$, and is a dash for
  the other seven. Its ``Additional costs'' column is populated once, for the \texttt{poly} DB-TFD row, with
  ``Shots for measuring $E$ and $V$''. ``Total Depth'' is not a column of its own, its header being
  fused with ``Assumptions/Remarks'' into a single cell whose entries are free text, so the table reports no
  depth for any row. There are no CNOT counts, no Toffoli counts and no circuit diagrams anywhere in the
  49 pages, and where the paper does use the phrase ``gate count'', in App.~G.3.4, it defines it as the
  query count ($N_g$ = Hamiltonian evolutions plus reflections). This is a query-complexity paper, which is
  precisely why the \texttt{oracle}/\texttt{pf} split exists.

One under-specification is resolved by a declared policy. Root ordering is not stated:
  the product is order-independent but the \emph{step sizes} are not, so the cost is not either. Measured on
  the tuning grid at $n = 4$, the five orderings we swept move the cost by at most $1.12\times$ at fixed
  Trotter angle budget (and not at all at $\beta = 0.2$). The $2.24\times$ spread across the full
  40-candidate grid at $\beta = 5$ is dominated by the angle budget, not by the ordering. Ordering is
  therefore a policy that must be declared, though its effect here is small. On the same grid the Chebyshev interpolant
  beats a truncated Taylor series in trace distance by $1.02\times$ at $\beta = 0.2$, $1.21$--$1.70\times$
  at $\beta = 1$ and $0.92$--$1.47\times$ at $\beta = 5$ (one ordering at $\beta = 5$ favors Taylor
  slightly), all at the tuning grid's single probe, $k = 3$ in the \texttt{pf} model. That probe is not
  where the frontier runs, so the two families were swept against each other to $k = 10$ in the
  \texttt{oracle} model at the tuned angle budget and root ordering
  (\texttt{run\_dbtfd.py --family-study}). The margin is strongly $k$-dependent and one-signed
  above $k = 1$: Taylor is marginally ahead at $k = 1$ ($1.06$ / $1.18$ / $1.01\times$ at
  $\beta = 0.2/1/5$), the two cross at $k = 2$, and at $k = 10$ Chebyshev is ahead by $253\times$,
  $556\times$ and $87\times$. The polynomial family is a small effect at the tuning probe and a
  two-to-three-order-of-magnitude effect at the iteration counts the frontier actually uses, which is a
  reason to declare it and not a reason to treat it as a detail.
  Our Chebyshev interpolant is taken on the \emph{exact} spectral
  interval, which is generous, since a real implementation would need a bound.

One cross-reference is misdirected. The main text says the Hamiltonian ``is
  normalized and shifted'' and that ``further details are provided in App.~G'', and App.~G contains no normalization or
  shift. The prescription does exist, in App.~E.2: $H \to aH - bI$ with $a = 2/\Delta E$ and
  $b = (E_{\max}+E_0)/\Delta E$. This is a mis-citation, not an omission, and the paper attaches its own
  caveat to the prescription (it ``requires knowledge of the maximum and minimum energies, which may not be
  feasible in practice''), together with the observation that loose spectral bounds shrink $a$ and lengthen
  the evolution. Both caveats anticipate this objection, and deviation D5 is exactly the shortcut they
  warn against.

Three observations point the other way. First, the polynomial is
named: App.~D is titled ``Polynomial approximation of exponential functions'', names Chebyshev polynomials
in its first paragraph and derives the $O(\sqrt{\tau})$-degree construction over two subsections. The
sentence about ``SciPy's root-finding functions'' is in App.~G.1 and concerns the factorization of an
already-chosen polynomial into linear factors so each can be mapped to a DB-QSP step. It is not the
specification of the polynomial family. The Chebyshev choice adopted here therefore follows the paper
rather than filling a gap in it. Second, the paper states the present open problem itself, in its own
discussion: ``establishing a comparison of query complexities within a common elementary gate set remains
an important and nontrivial goal.'' Third, the algorithm's ``no ancillas'' claim survives compilation, at
$12L-32$ CNOTs.

The paper's Fig.~4 plots
\emph{queries} against $\beta$, never the iteration count $k$. Inverting one into the other needs the
factor count $N_m$ of the product formula used, which the caption, the surrounding text and App.~G all
leave unstated, so no statement about the $k$ that figure implies can be made from the published text.
Its Lemma 2 is not comparable to a bare constant: it reads
$k \ge e\sqrt{(\beta/2)\log\bigl((8/\varepsilon_{\mathrm{DB}})\sqrt{D/Z(\beta)}\bigr)}$, i.e.\ a prefactor
$e/\sqrt2 \approx 1.92$ on $\sqrt{\beta\log(\cdot)}$ rather than $e$ on $\sqrt{\beta}$, with a logarithmic
factor that grows with $D$, with $1/\varepsilon$ and with $\beta$. Our measured
$k = (1.0\text{--}1.7)\sqrt{\bt}$ (Sec.~\ref{sec:clock}) carries no such factor and is a fit at
$\varepsilon = 0.05$ on one model family, so the two constants are not comparable.

\subsection{The local-driving Lindbladian protocol}
\label{sec:reimpl-ld}

Two versions of Ref.~\cite{hahn2025towards} exist and differ in respects material to the analysis
below. The version dates are recorded in the reference list. Unless stated otherwise, citations refer
to the second version (v2), and the first (v1) is cited explicitly where the versions differ.

\begin{figure}[t]
\centering
\includegraphics[width=0.50\textwidth]{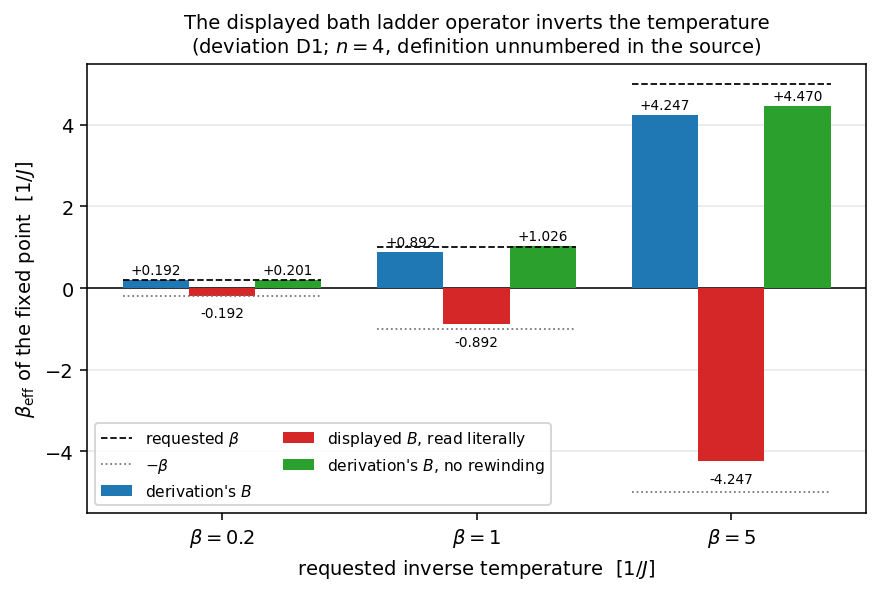}
\caption{The ladder-operator convention. Effective inverse temperature of the fixed point for the three
implemented variants at $n = 4$, at each $\beta$'s tuned filter width. Reading the source's displayed
definition literally ($B = \tfrac12(X_B - iY_B)$) inverts the sign of $\betaeff$ at every requested $\beta$:
the protocol converges toward $\rho_{-\beta}$. The inversion is exact rather than approximate, and it is
exact about the \emph{achieved} temperature and not the requested one: the literal reading returns
$-0.1917 / -0.8920 / -4.2467$ against the derivation-consistent $+0.1917 / +0.8920 / +4.2467$, matching to
every digit recorded, while the requested $\beta$ was $0.2 / 1 / 5$. Dashed and dotted rules mark $\beta$
and $-\beta$ so that the difference between the two statements is visible. The definition is unnumbered
inline prose in the source, in both versions, so no equation number is attached to it here.}
\label{fig:ladder}
\end{figure}

The first finding is an adjoint error in the bath ladder operator, present in both versions. The definition is
unnumbered inline prose in Sec.~III, in the sentence immediately after
the displayed system--bath Hamiltonian (v2 Eq.~(16)). It writes $B = \tfrac12(X_B - iY_B)$ while stating
in the same sentence that $B^\dagger|0\rangle = |1\rangle$ and $B|0\rangle = 0$. But
$\tfrac12(X - iY) = |1\rangle\langle 0|$, which \emph{raises}. The property the App.~B derivation actually
uses requires $B = |0\rangle\langle 1| = \tfrac12(X + iY)$. The requirement is essential to the derivation and the
printed formula is a typographical error. Its consequence for a reimplementer is maximal, because detailed balance in this protocol lives
entirely in the filter's imaginary shift $i\beta/4$ and the two readings differ by the sign of
$\mathrm{Im}\,f$: an implementation that adopts the printed formula converges to
$\rho_{-\beta}$. Measured
$\betaeff = -0.192 / -0.892 / -4.247$ for requested $\beta = 0.2/1/5$ at $n = 4$, against a
correct-convention $\betaeff$ of $0.1917 / 0.8920 / 4.2467$ at the same parameters: the literal reading
reproduces \emph{minus} the correct-convention value to four significant figures, with trace distances
0.41 / 0.97 / 1.00 (Fig.~\ref{fig:ladder}). The published text contains no indication of this
discrepancy. The source's own numerics (its Fig.~3) converge to positive $\beta$, which is consistent only with
the derivation's convention: this is a typographical slip in a displayed definition, not an error in its
results.
Reproduced by \texttt{ladder\_convention="literal"} and pinned as a test.

The version comparison sharpens this. The App.~B sentence that fixes the operator ordering was
reversed between versions. v1 writes that placing the bath in $|0\rangle$ ``forces the
application of $B_a(t_1)$ on the left and $B_a^\dagger(t_2)$ on the right'', whereas v2 writes
``$B_a^\dagger(t_1)$ on the left and $B_a(t_2)$ on the right''. The displayed equation that follows is
identical in content in both, though not in display or in number: v1's Eq.~(68) writes the superoperator
with a dot placeholder, $A_a(t_1)(\cdot)A_a^\dagger(t_2)$, and v2's Eq.~(75) writes $\rho$ explicitly. So
v1's stated ordering contradicts v1's own result line and v2's is the consistent one. The Sec.~III definition, however, is unchanged between versions: the correction went to the
derivation and not to the display it disagrees with.

The second finding is the Pareto dominance of the source's own $K'$ variant.
v1 reports, in its concluding remarks and as an observation it does not explain, that dropping
the rewinding step is \emph{more accurate}: ``A surprising preliminary result in this regard is that
the \emph{neglect} of the rewinding step, despite loosening the formal error bound, appears to improve the
convergence and accuracy of the resulting protocol.'' That sentence is deleted in v2, which instead
states the same finding twice: ``Surprisingly, our numerics indicate that removing the rewinding step
improves the performance of the protocol'' (Sec.~VI) and ``\ldots the accuracy of these results improves
in comparison to the protocol $K[\rho]$. Understanding the reason for this improvement remains an open
question for future work'' (Sec.~V, where the source sentence opens ``It turns out that''). Under a gate count the statement becomes much stronger, because
rewinding is \emph{exactly} the extra $T$ of Hamiltonian evolution per cycle: at a matched filter width
$K'$ is $1.44$--$1.67\times$ cheaper \emph{and} $2.7$--$19.4\times$ more accurate at all nine
$(n,\beta)$ cells (Table~\ref{tab:krewind}). The accuracy component is the paper's own observation, in
both versions. The cost component arises only under an explicit gate count.

The third finding is the suboptimality of the filter width used in the source numerics. The source's
numerical case study runs at $\sigma = 0.5$ (v2 Sec.~V and Fig.~3, whereas Fig.~2 is a schematic of the chain of
approximations, not a numerical figure), against the $\sigma^\star$ of Sec.~\ref{sec:floors}. That is a
confirmation of the paper's own prediction rather than an oversight, and here the version matters:
v2 Sec.~IV~D, ``Optimal Accuracy'', predicts exactly this (``one thus expects the mixing time to
diverge \emph{exponentially}'' and ``there exists a finite $\sigma^*$ which minimizes the fixed-point
error''), and its App.~E measures it. v1 asserts the opposite: it has no Sec.~IV~D and no
App.~E, and its Sec.~IV~C closes with the statement ``We do not expect unfavorable scaling with $\sigma$
in the general case''. The measurement is therefore a confirmation of v2 and a counterexample to v1, and
it is why $\sigma$ is tuned per $\beta$ here.

The fourth finding is that the published parameters lie outside the analyzed regime. v2 states this directly:
``We note that we are in a regime $\beta \gtrsim \sigma$ where the error bounds in our protocol are not
necessarily optimal'' (Sec.~V). The bound degrades further: v2's Magnus bound carries a prefactor
$e^{\beta^2/2\sigma^2}$ (Eqs.~(20) and (30)), which is $e^2$ at the case study's parameters and
$e^{50} \approx 5\times10^{21}$ at $\beta = 5$, $\sigma = 0.5$. The guarantee is vacuous precisely in
the regime a benchmark must occupy. Neither the qualification nor the prefactor exists in v1, whose App.~B
instead uses $\beta \lesssim \sigma$ as a premise of the proof (``Elsewhere in the proof, we have that
$\beta \lesssim \sigma$''), a sentence deleted in v2. The objection therefore applies to v1, while v2's
own qualification anticipates it. This is the second of three arms whose published numerics fall
outside the asymptotic regime their theorems address, here as the source itself notes.

The fifth finding is the absence of an ambiguity in the time window. The
time window is the natural candidate for an inconsistency in this protocol, because the
rewinding step doubles a duration and the interaction picture obscures where that doubling is applied.
No inconsistency is present. Searching both versions exhaustively, every integration limit is
$\int_{-T/2}^{T/2}$ without exception, the convention is stated once in prose (``We assume that the
$f(t)$ remains operational over some finite time interval $[-T/2, T/2]$, chosen to be symmetric for
convenience''), and the only place a bare $T$ appears as a duration is the cycle length, which is the
width of that same window. Our $K'$ is implemented as dropping the rewinding block, which is what the
paper describes, and deviation D5 is retained only to record that the interaction-picture telescoping in
App.~\ref{app:protocols} makes the rewind block explicit rather than reading it off the formula.

The sixth finding is an unstated ergodicity condition on the published jump family. App.~E (v2) takes
$A_i = Y_i$, one Pauli letter translated along the chain, and neither version states a condition on the
Hamiltonian for the resulting sampler to have a \emph{unique} fixed point. There is one, and it is
restrictive: $S_Y = \sum_i Y_i$ commutes with every $A_a$ by construction, so on any $H$ commuting with
$S_Y$ the exact KMS-detailed-balanced sampler acquires one stationary state per $S_Y$ eigenvalue.
Measured on the isotropic Heisenberg chain at $n = 4$: a five-dimensional kernel, and a protocol floor
of $0.5009$ that the sector-projection argument predicts to four decimals
(Sec.~\ref{sec:heisenberg}). The paper's own numerics use a mixed-field Ising model, where the condition
is satisfied, so nothing it reports is affected. An implementation using a different model receives no
warning from the published text.
Pinned as a test.

\begin{table}[t]
\centering\small
\caption{Rewound $K$ against no-rewinding $K'$ at matched filter width. $K'$ Pareto-dominates $K$ at
every cell. Both channels are evaluated at one matched filter
width per row, namely $K$'s own $\sigma^\star$ from Table~\ref{tab:floors}, so that the cost and accuracy
ratios are ratios at the same $\sigma$ and not a comparison of two differently-tuned protocols. That choice
is the least favorable one for the $K'$ claim, since it gives $K$ its best width and $K'$ a width that is
not always its best. The three rows where $K'$ would do better at its own $\sigma^\star$ are marked
$\ddagger$ and its value there is given in the text. Floors are final values, as in
Table~\ref{tab:floors}. The Hamiltonian-time ratio is exactly 0.5 by construction. The gate ratio is
0.60--0.70 because the drive pulses are not halved. Every row is a converged fixed point (last-decile
drift at or below 1\,\%) except the $n = 8$, $\beta = 0.2$ row, whose $K$ column is still descending at
that size's 300-cycle horizon (Table~\ref{tab:floors}).}
\label{tab:krewind}
\begin{tabular}{@{}lllrrrrrrr@{}}
\toprule
$n$ & $\beta$ & $\bt$ & $\sigma$ & $K$ \twoq/cyc & $K'$ \twoq/cyc & cost ratio & $K$ floor & $K'$ floor & accuracy ratio \\
\midrule
4 & 0.2 &  0.95 & 0.75 & 1256 & 872 & 0.694 & 0.0286 & 0.0029$^\ddagger$ & 10.0$\times$ \\
4 & 1 &  4.76 & 2 & 3320 & 2312 & 0.696 & 0.0578 & 0.0199$^\ddagger$ & 2.9$\times$ \\
4 & 5 & 23.79 & 3 & 7804 & 4804 & 0.616 & 0.0143 & 0.0024 & 6.0$\times$ \\
6 & 0.2 &  1.46 & 1.5 & 6288 & 4328 & 0.688 & 0.0305 & 0.0016 & 19.4$\times$ \\
6 & 1 &  7.30 & 1.5 & 6288 & 4328 & 0.688 & 0.1067 & 0.0158 & 6.7$\times$ \\
6 & 5 & 36.48 & 3 & 19916 & 12036 & 0.604 & 0.0663 & 0.0052 & 12.8$\times$ \\
8 & 0.2 &  1.97 & 1 & 7936 & 5416 & 0.682 & 0.0391 & 0.0115$^\ddagger$ & 3.4$\times$ \\
8 & 1 &  9.84 & 1.5 & 11832 & 8080 & 0.683 & 0.1419 & 0.0528 & 2.7$\times$ \\
8 & 5 & 49.19 & 3 & 37580 & 22516 & 0.599 & 0.1115 & 0.0156 & 7.1$\times$ \\
\bottomrule
\end{tabular}
\end{table}

\noindent At the three $\ddagger$ rows $K'$ does better at its own $\sigma^\star$. At $n = 4$, $\beta = 0.2$
it reaches 0.0014 at $\sigma = 1$ rather than 0.0029. At $n = 4$, $\beta = 1$ it reaches 0.0176 at
$\sigma = 1.5$ rather than 0.0199. At $n = 8$, $\beta = 0.2$ it reaches 0.0084 at $\sigma = 0.75$ rather
than 0.0115. Using those instead would raise the three accuracy ratios to $20.1\times$, $3.3\times$ and
$4.7\times$. Table~\ref{tab:krewind} prints the matched-$\sigma$ figures because a ratio between two differently-tuned
configurations is not a ratio between two protocols. Across the table the cost ratio is 0.599--0.696, i.e.\
$K'$ is $1.44$--$1.67\times$ cheaper per cycle, and the accuracy ratio is $2.7$--$19.4\times$. Both ends of
the accuracy range are converged cells.

\section{Discussion}
\label{sec:discussion}

The recommendation that follows from Sec.~\ref{sec:results} is procedural rather than algorithmic: a
frontier should be reported per arm, with the cost model, the side channels and the tuning budget
declared, and the compilation regime reported as an axis, in place of a single winner. Under that
discipline the identity of the best method here moved on three of those four axes on the Ising grid, and
on the fourth, the axis declared in advance as the one most likely to matter, only once a second model
family was added (Sec.~\ref{sec:heisenberg}) or the matching metric changed to free energy
(Sec.~\ref{sec:sensitivity}). An axis that moves nothing in 36 cells and one thing in the next eight is an
argument for reporting it rather than for dropping it.

The comparison covers two model families and one tuning-budget currency. The critical 1D TFIM carries the whole grid, and the Heisenberg cross-check of
Sec.~\ref{sec:heisenberg} adds $n = 4, 6$ at $\beta = 1$. Two families are enough to show that the floor
taxonomy is not a free-fermion artifact, to add a fourth floor mechanism and to produce the one
compilation-regime inversion a single family could not. They leave the inversion \emph{counts} a property
of this grid alone, and two of the three move in opposite directions between the two models. The currency
is likewise a choice. Exact-reference evaluations are the one currency all three arms spend, but
they are neither wall-clock nor shots, so an arm with a cheap inner loop and an expensive outer loop is
charged the same as its opposite. The 200 calls per $\beta$ were fixed before any result.
Sec.~\ref{sec:budgetfrontier} prices the gap between spending that budget and spending the whole swept
grid, and every headline count is reported under both.

Every arm is simulated exactly, by CPU density matrix or Kraus map, with no shot noise, no device noise
and no routing, so the absolute gate counts are lower bounds on hardware. Shot noise is not simulated for
DB-TFD, whose own Fig.~3 shows the error to be product-formula-dominated above $10^5$ shots, and the other
two arms have none to simulate. The measurement load enters as a declared side channel
(Sec.~\ref{sec:sidechannels}) instead.

Two conclusions should transfer beyond the three protocols compared here. First, the gap between an
algorithm's own resource currency and a realizable circuit is not a constant
(Sec.~\ref{sec:oraclegap}): it is anticorrelated with
how analog-native the algorithm is, and it is large enough to reorder rankings. That the gap exists is
known~\cite{babbush2021focus,hoefler2023disentangling,beverland2022assessing,dalzell2025quantum,jaques2025qram}.
That its size is a property of the algorithm family rather than a constant overhead is what a
common-gate-set comparison adds, and it is what makes a ranking assembled from published resource
statements biased with a known sign rather than merely imprecise. Its sign and its order of magnitude
survive both model families measured here, and the same behavior is expected wherever a declared cost
model and a compiled circuit stand in that relation. Second, all three
source papers carried a discrepancy that reading alone did not surface, and two of the three publish
numerics outside the asymptotic regime their theorems address
(Sec.~\ref{sec:reimpl}). Both observations
concern practice across the field rather than any individual paper, and both are inexpensive to address
with a shared harness to which further arms can be added. A new arm requires only a
\texttt{(term, time)} schedule and a declaration of its cost model. The same process located one error in
the harness itself, recorded in App.~\ref{sec:ourbug}, whose guard is the general one: a benchmark that
tests only counts cannot detect a dynamics error, so the counted circuit must also be simulated.

The remaining findings extend less far. The floor taxonomy is a statement about these three protocols,
now checked on two models and extended by the second, and the inversion counts are properties of a grid.
One cell has been carried a size further: under the envelope policy the three-leader sequence persists at
$n = 9$, $\beta = 5$, with the $10^4$ margin shrinking to $1.03\times$ (Sec.~\ref{sec:inversions}).
The next test of the taxonomy is therefore a fourth arm rather than a third model.
Two additions would sharpen the frontier most, in order of cost. A fourth and fifth arm, the
block-encoding-free local-circuit sampler of Ref.~\cite{hahn2026efficient} and a QITE-family
method~\cite{motta2020determining,tang2026interference}, would test the oracle-gap claim on families whose
analog-nativeness sits between the two extremes measured here. The counting itself can be delegated:
Ref.~\cite{harrigan2024qualtran} is a better costing backend than ours and would extend these tables to
Clifford$+T$ and to surface-code space-time volume, provided the counted-equals-simulated guarantee of
Sec.~\ref{sec:accounting} is kept as a separate check on top of it, since a resource count alone does not
say whether the algorithm reached the state.

\section*{Data and code availability}
\addcontentsline{toc}{section}{Data and code availability}

The harness, the raw simulation artifacts and the analysis scripts are kept together in one git
repository, which also holds this manuscript's source. The code, data and run logs are available from the
author on request. In that repository, every number reported above is reproduced by
the command sequence in its \texttt{README.md} from a clean checkout, CPU only, under Python
3.12 with \texttt{numpy}, \texttt{scipy} and \texttt{matplotlib} at the versions pinned in
\texttt{requirements.txt}, with no build step. The validation suite is 186 fast tests plus 6 marked slow,
all passing at the commit these numbers were produced from. Every finding in Sec.~\ref{sec:reimpl} is
pinned by a named test, and the two findings concerning Eq.~(G9) of Ref.~\cite{wright2026double} are pinned
as a pair, one test asserting that the formula as printed does not reach fourth order and a second that
the formula in the source it cites does.

Raw artifacts (\texttt{data/*/\*.json}) carry, for every run, every configuration field, both compilation
regimes, both cost models, the declared side channels, the spectral scale $\lambda$, $\bt = \beta\lambda$,
$t^\star$ where it is defined, \texttt{cumulative\_hamiltonian\_time}, \texttt{effective\_beta}, the
last-decile drift, and the boolean \texttt{circuit\_is\_the\_simulation} required by N1.
\texttt{data/MANIFEST.md} maps every table and figure above to the file and field it reads,
\texttt{data/README.md} documents the six single-$\beta$ partial sweeps and the concatenation that
produced the merged $n = 6, 8$ Lindbladian artifacts, and \texttt{scripts/backfill\_artifacts.py --check}
verifies that no artifact is missing a documented field. The simulation artifacts under \texttt{data/}
are those of the repository's commit \texttt{9e0f33e}, the state every number above was produced from.

\bibliographystyle{unsrt}
\bibliography{refs}

\begin{thebibliography}{10}

\bibitem{ding2025simple}
Zhiyan Ding, Yongtao Zhan, John Preskill, and Lin Lin.
\newblock Simple and efficient end-to-end quantum thermal and ground state
  preparation.
\newblock {\em Nature Physics}, 2026.
\newblock arXiv:2508.05703 (v1: 7 Aug 2025; v2: 18 Dec 2025; v3: 12 Aug 2026);
  published online 12 August 2026; statements cited here are from v3.

\bibitem{ding2024single}
Zhiyan Ding, Chi-Fang Chen, and Lin Lin.
\newblock Single-ancilla ground state preparation via {L}indbladians.
\newblock {\em Physical Review Research}, 6:033147, 2024.

\bibitem{farrell2026preparing}
Roland~C. Farrell, Yongtao Zhan, Lucas Katschke, Lode Pollet, Ilan~T. Rosen,
  and Jad~C. Halimeh.
\newblock Preparing thermal states of frustrated quantum spin systems using 139
  qubits, 2026.
\newblock arXiv:2605.26245 (v1: 25 May 2026; v2: 29 June 2026); 32 pages, 21
  figures. v2 extends the lattices from 24 to 27 sites; the Methods D
  statements cited here are identical in both versions.

\bibitem{wright2026double}
Andrew Wright, Reyhaneh Aghaei~Saem, Supanut Thanasilp, Yudai Suzuki, and
  Zo\"{e} Holmes.
\newblock Double-bracket quantum algorithms for thermal state preparation,
  2026.
\newblock arXiv:2606.05947 (v1: 4 June 2026, the only version); 49 pages.

\bibitem{gluza2026double}
Marek Gluza, Jeongrak Son, Bi~Hong Tiang, Ren\'{e} Zander, Raphael Seidel,
  Yudai Suzuki, Zo\"{e} Holmes, and Nelly H.~Y. Ng.
\newblock Double-bracket quantum algorithms for quantum imaginary-time
  evolution.
\newblock {\em Physical Review Letters}, 136:020601, 2026.
\newblock arXiv:2412.04554.

\bibitem{wu2019variational}
Jingxiang Wu and Timothy~H. Hsieh.
\newblock Variational thermal quantum simulation via thermofield double states.
\newblock {\em Physical Review Letters}, 123:220502, 2019.

\bibitem{hahn2025towards}
Dominik Hahn, S.~A. Parameswaran, and Benedikt Placke.
\newblock Towards efficient quantum thermal state preparation via local
  driving: Lindbladian simulation with provable guarantees, 2025.
\newblock arXiv:2505.22816 (v1: 28 May 2025, 28 pages, 3 figures; v2: 13
  January 2026, 31 pages, 4 figures).

\bibitem{chen2023quantum}
Chi-Fang Chen, Michael~J. Kastoryano, Fernando G. S.~L. Brand\~{a}o, and
  Andr\'{a}s Gily\'{e}n.
\newblock Quantum thermal state preparation, 2023.
\newblock arXiv:2303.18224 (v1: 31 March 2023); 79 pages, 12 figures.

\bibitem{chen2023efficient}
Chi-Fang Chen, Michael~J. Kastoryano, and Andr\'{a}s Gily\'{e}n.
\newblock An efficient and exact noncommutative quantum {G}ibbs sampler, 2023.
\newblock arXiv:2311.09207 (v1: 15 November 2023; v2: 14 October 2025);
  preprint.

\bibitem{ding2025efficient}
Zhiyan Ding, Bowen Li, and Lin Lin.
\newblock Efficient quantum {G}ibbs samplers with {K}ubo--{M}artin--{S}chwinger
  detailed balance condition.
\newblock {\em Communications in Mathematical Physics}, 406:67, 2025.
\newblock arXiv:2404.05998 (v1: 9 April 2024; v5: 29 October 2024).

\bibitem{rouze2026optimal}
Cambyse Rouz\'{e}, Daniel Stilck~Fran\c{c}a, and \'{A}lvaro~M. Alhambra.
\newblock Optimal quantum algorithm for {G}ibbs state preparation.
\newblock {\em Physical Review Letters}, 136:060601, 2026.
\newblock arXiv:2411.04885 (v1: 7 November 2024; v2: 10 February 2026).

\bibitem{rouze2026efficient}
Cambyse Rouz\'{e}, Daniel Stilck~Fran\c{c}a, and \'{A}lvaro~M. Alhambra.
\newblock Efficient thermalization and universal quantum computing with quantum
  {G}ibbs samplers.
\newblock {\em Nature Physics}, 22(7):1123--1127, 2026.
\newblock arXiv:2403.12691 (v1: 19 March 2024; v3: 17 April 2026).

\bibitem{brunner2025lindblad}
Eric Brunner, Luuk Coopmans, Gabriel Matos, Matthias Rosenkranz, Frederic
  Sauvage, and Yuta Kikuchi.
\newblock Lindblad engineering for quantum {G}ibbs state preparation under the
  eigenstate thermalization hypothesis.
\newblock {\em Quantum}, 9:1843, 2025.
\newblock arXiv:2412.17706; 62 pages, 22 figures.

\bibitem{chen2025randomized}
Hongrui Chen, Bowen Li, Jianfeng Lu, and Lexing Ying.
\newblock A randomized method for simulating {L}indblad equations and thermal
  state preparation.
\newblock {\em Quantum}, 9:1917, 2025.
\newblock arXiv:2407.06594.

\bibitem{lloyd2025quantum}
Jerome Lloyd and Dmitry~A. Abanin.
\newblock Quantum thermal state preparation for near-term quantum processors,
  2025.
\newblock arXiv:2506.21318 (v1: 26 June 2025).

\bibitem{rall2023thermal}
Patrick Rall, Chunhao Wang, and Pawel Wocjan.
\newblock Thermal state preparation via rounding promises.
\newblock {\em Quantum}, 7:1132, 2023.
\newblock arXiv:2210.01670 (v1: 4 October 2022; v2: 4 October 2023).

\bibitem{motta2020determining}
Mario Motta, Chong Sun, Adrian T.~K. Tan, Matthew~J. O'Rourke, Erika Ye,
  Austin~J. Minnich, Fernando G. S.~L. Brand\~{a}o, and Garnet Kin-Lic Chan.
\newblock Determining eigenstates and thermal states on a quantum computer
  using quantum imaginary time evolution.
\newblock {\em Nature Physics}, 16:205--210, 2020.
\newblock arXiv:1901.07653 (v1: 23 January 2019; v3: 16 February 2020).

\bibitem{tang2026interference}
Hong-Jian Tang and Dan-Bo Zhang.
\newblock Interference engineering for quantum imaginary-time evolution through
  multiple energy shifts, 2026.
\newblock arXiv:2608.17792 (v1: 18 August 2026); 7 pages, 5 figures.

\bibitem{ramkumar2024mixing}
Akshar Ramkumar and Mehdi Soleimanifar.
\newblock Mixing time of quantum {G}ibbs sampling for random sparse
  {H}amiltonians, 2024.
\newblock arXiv:2411.04454 (v1: 7 November 2024); 31 pages, 1 figure.

\bibitem{temme2011quantum}
K.~Temme, T.~J. Osborne, K.~G. Vollbrecht, D.~Poulin, and F.~Verstraete.
\newblock Quantum {M}etropolis sampling.
\newblock {\em Nature}, 471:87, 2011.

\bibitem{yung2012quantum}
Man-Hong Yung and Al{\'a}n Aspuru-Guzik.
\newblock A quantum-quantum {M}etropolis algorithm.
\newblock {\em Proceedings of the National Academy of Sciences}, 109:754--759,
  2012.

\bibitem{poulin2009sampling}
David Poulin and Pawel Wocjan.
\newblock Sampling from the thermal quantum {G}ibbs state and evaluating
  partition functions with a quantum computer.
\newblock {\em Physical Review Letters}, 103:220502, 2009.

\bibitem{chowdhury2017quantum}
Anirban~Narayan Chowdhury and Rolando~D. Somma.
\newblock Quantum algorithms for {G}ibbs sampling and hitting-time estimation.
\newblock {\em Quantum Information and Computation}, 17(1--2):41--64, 2017.

\bibitem{wocjan2023szegedy}
Pawel Wocjan and Kristan Temme.
\newblock Szegedy walk unitaries for quantum maps.
\newblock {\em Communications in Mathematical Physics}, 402(3):3201--3231,
  2023.

\bibitem{cleve2017efficient}
Richard Cleve and Chunhao Wang.
\newblock Efficient quantum algorithms for simulating {L}indblad evolution.
\newblock In {\em 44th International Colloquium on Automata, Languages, and
  Programming (ICALP 2017)}, volume~80 of {\em LIPIcs}, pages 17:1--17:14,
  2017.

\bibitem{bergamaschi2024quantum}
Thiago Bergamaschi, Chi-Fang Chen, and Yunchao Liu.
\newblock Quantum computational advantage with constant-temperature {G}ibbs
  sampling.
\newblock In {\em 2024 IEEE 65th Annual Symposium on Foundations of Computer
  Science (FOCS)}, 2024.

\bibitem{verdon2019quantum}
Guillaume Verdon, Jacob Marks, Sasha Nanda, Stefan Leichenauer, and Jack
  Hidary.
\newblock Quantum {H}amiltonian-based models and the variational quantum
  thermalizer algorithm, 2019.

\bibitem{chowdhury2020variational}
Anirban~N. Chowdhury, Guang~Hao Low, and Nathan Wiebe.
\newblock A variational quantum algorithm for preparing quantum {G}ibbs states,
  2020.

\bibitem{sagastizabal2021variational}
R.~Sagastizabal, S.~P. Premaratne, B.~A. Klaver, M.~A. Rol, V.~Neg{\^\i}rneac,
  M.~Moreira, X.~Zou, S.~Johri, N.~Muthusubramanian, M.~Beekman,
  C.~Zachariadis, V.~P. Ostroukh, N.~Haider, A.~Bruno, A.~Y. Matsuura, and
  L.~DiCarlo.
\newblock Variational preparation of finite-temperature states on a quantum
  computer.
\newblock {\em npj Quantum Information}, 7:130, 2021.

\bibitem{selisko2024extending}
Johannes Selisko, Maximilian Amsler, Thomas Hammerschmidt, Ralf Drautz, and
  Thomas Eckl.
\newblock Extending the variational quantum eigensolver to finite temperatures.
\newblock {\em Quantum Science and Technology}, 9(1):015026, 2024.

\bibitem{hahn2026efficient}
Dominik Hahn, Ryan Sweke, Abhinav Deshpande, and Oles Shtanko.
\newblock Efficient quantum {G}ibbs sampling with local circuits.
\newblock {\em PRX Quantum}, 7:020314, 2026.
\newblock arXiv:2506.04321 (v1: 4 June 2025; v2: 13 May 2026).

\bibitem{babbush2021focus}
Ryan Babbush, Jarrod~R. McClean, Michael Newman, Craig Gidney, Sergio Boixo,
  and Hartmut Neven.
\newblock Focus beyond quadratic speedups for error-corrected quantum
  advantage.
\newblock {\em PRX Quantum}, 2:010103, 2021.
\newblock arXiv:2011.04149 (v1: 9 November 2020; v2: 31 March 2021).

\bibitem{hoefler2023disentangling}
Torsten Hoefler, Thomas H\"{a}ner, and Matthias Troyer.
\newblock Disentangling hype from practicality: on realistically achieving
  quantum advantage.
\newblock {\em Communications of the ACM}, 66(5):82--87, 2023.
\newblock arXiv:2307.00523 (v1: 2 July 2023).

\bibitem{beverland2022assessing}
Michael~E. Beverland, Prakash Murali, Matthias Troyer, Krysta~M. Svore, Torsten
  Hoefler, Vadym Kliuchnikov, Guang~Hao Low, Mathias Soeken, Aarthi Sundaram,
  and Alexander Vaschillo.
\newblock Assessing requirements to scale to practical quantum advantage, 2022.
\newblock arXiv:2211.07629 (v1: 14 November 2022).

\bibitem{dalzell2025quantum}
Alexander~M. Dalzell, Sam McArdle, Mario Berta, Przemyslaw Bienias, Chi-Fang
  Chen, Andr\'{a}s Gily\'{e}n, Connor~T. Hann, Michael~J. Kastoryano, Emil~T.
  Khabiboulline, Aleksander Kubica, Grant Salton, Samson Wang, and Fernando G.
  S.~L. Brand\~{a}o.
\newblock {\em Quantum Algorithms: A Survey of Applications and End-to-End
  Complexities}.
\newblock Cambridge University Press, 2025.
\newblock arXiv:2310.03011 (v1: 4 October 2023; v2: 5 August 2025).

\bibitem{sanders2020compilation}
Yuval~R. Sanders, Dominic~W. Berry, Pedro C.~S. Costa, Louis~W. Tessler, Nathan
  Wiebe, Craig Gidney, Hartmut Neven, and Ryan Babbush.
\newblock Compilation of fault-tolerant quantum heuristics for combinatorial
  optimization.
\newblock {\em PRX Quantum}, 1:020312, 2020.
\newblock arXiv:2007.07391 (v1: 14 July 2020; v2: 5 August 2020).

\bibitem{jaques2025qram}
Samuel Jaques and Arthur~G. Rattew.
\newblock {QRAM}: a survey and critique.
\newblock {\em Quantum}, 9:1922, 2025.
\newblock arXiv:2305.10310 (v1: 17 May 2023; v2: 9 November 2025).

\bibitem{liu2026logical}
Danylo Liu, Hanrui Wang, Shuo Zhu, Jason Cong, Wibe~A. de~Jong, Di~Fang, Zhiyan
  Huang, Costin Iancu, and Chao Yang.
\newblock Logical resource estimation for quantum state preparation with
  compilation, 2026.
\newblock arXiv:2605.18877 (v1: 15 May 2026).

\bibitem{bonetmonroig2023performance}
Xavier Bonet-Monroig, Hao Wang, Diederick Vermetten, Bruno Senjean, Charles
  Moussa, Thomas B\"{a}ck, Vedran Dunjko, and Thomas~E. O'Brien.
\newblock Performance comparison of optimization methods on variational quantum
  algorithms.
\newblock {\em Physical Review A}, 107:032407, 2023.
\newblock arXiv:2111.13454 (v1: 26 November 2021; v3: 8 March 2023).

\bibitem{proctor2025benchmarking}
Timothy Proctor, Kevin Young, Andrew~D. Baczewski, and Robin Blume-Kohout.
\newblock Benchmarking quantum computers.
\newblock {\em Nature Reviews Physics}, 7:105--118, 2025.
\newblock arXiv:2407.08828 (v1: 11 July 2024; v2: 13 February 2026).

\bibitem{bowles2024better}
Joseph Bowles, Shahnawaz Ahmed, and Maria Schuld.
\newblock Better than classical? the subtle art of benchmarking quantum machine
  learning models, 2024.
\newblock arXiv:2403.07059 (v1: 11 March 2024; v2: 14 March 2024); preprint.

\bibitem{lucic2018gans}
Mario Lucic, Karol Kurach, Marcin Michalski, Sylvain Gelly, and Olivier
  Bousquet.
\newblock Are {GAN}s created equal? a large-scale study.
\newblock In {\em Advances in Neural Information Processing Systems 31}, pages
  698--707, 2018.
\newblock arXiv:1711.10337 (v1: 28 November 2017; v4: 29 October 2018).

\bibitem{melis2018state}
G\'{a}bor Melis, Chris Dyer, and Phil Blunsom.
\newblock On the state of the art of evaluation in neural language models.
\newblock In {\em International Conference on Learning Representations (ICLR)},
  2018.
\newblock arXiv:1707.05589 (v1: 18 July 2017; v2: 20 November 2017).

\bibitem{ferraridacrema2019progress}
Maurizio Ferrari~Dacrema, Paolo Cremonesi, and Dietmar Jannach.
\newblock Are we really making much progress? a worrying analysis of recent
  neural recommendation approaches.
\newblock In {\em Proceedings of the 13th ACM Conference on Recommender Systems
  (RecSys)}, pages 101--109, 2019.
\newblock arXiv:1907.06902 (v1: 16 July 2019; v3: 16 August 2019).

\bibitem{dodge2019show}
Jesse Dodge, Suchin Gururangan, Dallas Card, Roy Schwartz, and Noah~A. Smith.
\newblock Show your work: improved reporting of experimental results.
\newblock In {\em Proceedings of EMNLP-IJCNLP 2019}, pages 2185--2194, 2019.
\newblock arXiv:1909.03004 (v1: 6 September 2019).

\bibitem{tang2020showing}
Raphael Tang, Jaejun Lee, Ji~Xin, Xinyu Liu, Yaoliang Yu, and Jimmy Lin.
\newblock Showing your work doesn't always work.
\newblock In {\em Proceedings of ACL 2020}, pages 2766--2772, 2020.
\newblock arXiv:2004.13705 (v1: 28 April 2020).

\bibitem{bouthillier2021accounting}
Xavier Bouthillier, Pierre Delaunay, Mirko Bronzi, Assya Trofimov, Brennan
  Nichyporuk, Justin Szeto, Nazanin Sepah, Edward Raff, Kanika Madan, Vikram
  Voleti, Samira Ebrahimi~Kahou, Vincent Michalski, Dmitriy Serdyuk, Tal Arbel,
  Chris Pal, Ga\"{e}l Varoquaux, and Pascal Vincent.
\newblock Accounting for variance in machine learning benchmarks.
\newblock In {\em Proceedings of Machine Learning and Systems (MLSys)},
  volume~3, 2021.
\newblock arXiv:2103.03098 (v1: 1 March 2021).

\bibitem{lubinski2023application}
Thomas Lubinski, Sonika Johri, Paul Varosy, Jeremiah Coleman, Luning Zhao,
  Jason Necaise, Charles~H. Baldwin, Karl Mayer, and Timothy Proctor.
\newblock Application-oriented performance benchmarks for quantum computing.
\newblock {\em IEEE Transactions on Quantum Engineering}, 4:1--32, 2023.
\newblock arXiv:2110.03137 (v1: 7 October 2021; v3: 10 January 2023).

\bibitem{chatterjee2025comprehensive}
Avimita Chatterjee, Sonny Rappaport, Anish Giri, Sonika Johri, Timothy Proctor,
  David~E. Bernal~Neira, Pratik Sathe, and Thomas Lubinski.
\newblock A comprehensive cross-model framework for benchmarking the
  performance of quantum {H}amiltonian simulations.
\newblock {\em IEEE Transactions on Quantum Engineering}, 6:1--26, 2025.
\newblock arXiv:2409.06919 (v1: 11 September 2024).

\bibitem{harrigan2024qualtran}
Matthew~P. Harrigan, Tanuj Khattar, Charles Yuan, Anurudh Peduri, Noureldin
  Yosri, Fionn~D. Malone, Ryan Babbush, and Nicholas~C. Rubin.
\newblock Expressing and analyzing quantum algorithms with {Qualtran}, 2024.
\newblock arXiv:2409.04643 (v1: 6 September 2024).

\bibitem{sharma2024quantum}
Arjun Sharma and Kevin Obenland.
\newblock Quantum resources for pure thermal shadows, 2024.
\newblock arXiv:2409.05777 (v1: 9 September 2024); IEEE QCE24 proceedings.

\bibitem{wu2024variational}
Dian Wu, Riccardo Rossi, Filippo Vicentini, et~al.
\newblock Variational benchmarks for quantum many-body problems.
\newblock {\em Science}, 386(6719):296--301, 2024.
\newblock arXiv:2302.04919.

\bibitem{fawzi2024certified}
Hamza Fawzi, Omar Fawzi, and Samuel~O. Scalet.
\newblock Certified algorithms for equilibrium states of local quantum
  {H}amiltonians.
\newblock {\em Nature Communications}, 15:7394, 2024.
\newblock arXiv:2311.18706 (v1: 30 November 2023; v3: 14 August 2025).

\bibitem{granet2026adiabatic}
Etienne Granet and Henrik Dreyer.
\newblock Adiabatic preparation of thermal states and entropy--noise relation
  on noisy quantum computers.
\newblock {\em npj Quantum Information}, 12:105, 2026.
\newblock arXiv:2509.05206 (v1: 5 September 2025; v2: 20 April 2026).

\bibitem{white2009minimally}
Steven~R. White.
\newblock Minimally entangled typical quantum states at finite temperature.
\newblock {\em Physical Review Letters}, 102:190601, 2009.

\bibitem{stoudenmire2010minimally}
E.~M. Stoudenmire and Steven~R. White.
\newblock Minimally entangled typical thermal state algorithms.
\newblock {\em New Journal of Physics}, 12:055026, 2010.

\bibitem{sugiura2012thermal}
Sho Sugiura and Akira Shimizu.
\newblock Thermal pure quantum states at finite temperature.
\newblock {\em Physical Review Letters}, 108:240401, 2012.

\bibitem{chen2022efficient}
Yu-An Chen, Andrew~M. Childs, Mohammad Hafezi, Zhang Jiang, Hwanmun Kim, and
  Yijia Xu.
\newblock Efficient product formulas for commutators and applications to
  quantum simulation.
\newblock {\em Physical Review Research}, 4:013191, 2022.
\newblock arXiv:2111.12177 (v1: 23 November 2021).

\end{thebibliography}

\appendix
\small

\section{Protocol implementations}
\label{app:protocols}

\subsection{The single-ancilla system--bath protocol (arm 1)}

Sources: Ref.~\cite{ding2025simple} (algorithm, fixed-point theorem) and Ref.~\cite{farrell2026preparing}
(all numerical hyperparameters). One reset cycle: (i) draw $\omega \sim \mathrm{Uniform}(0,\omega_{\max}]$,
a site $i_S$, and a jump operator $O_S \in \{X,Y,Z\}$ at that site, all uniformly and all redrawn every
cycle as in the source papers, (ii) prepare the ancilla in the Gibbs state of
$H_E = -\tfrac12\omega Z_E$, i.e.\ $\rho_E = \mathrm{diag}(1-p,p)$ with $p(\omega) = 1/(1+e^{\beta\omega})$,
(iii) evolve system plus ancilla over $t \in [-T/2, T/2]$ under $H(t) = H_S + H_E + \alpha f(t)H_{SE}$ with
$H_{SE} = O_S \otimes X_E$, Trotterized into $M = T/\delta t$ steps, step $j$ freezing the envelope at
$t_j = -T/2 + (j+\tfrac12)\delta t$, and (iv) reset the ancilla to $|0\rangle$.
The envelope is $f(t) = N\exp[-t^2/(4\sigma^2T^2)]$ with $\sum_j \delta t\,f(t_j)^2 = 1$, the discrete form
of Ref.~\cite{ding2025simple}'s $\int f^2 = 1$. Here $\sigma$ is dimensionless, the physical width is $\sigma T$,
and the spectral width detailed balance cares about is $1/(\sigma T)$. $H_{SE} = O_S \otimes X_E$ follows
from Eq.~(1) of Ref.~\cite{ding2025simple} with $B_E = |1\rangle\langle 0|$, which is that paper's own
choice, and with $A_S$ specialized \emph{by us} to a Hermitian Pauli. This narrows the source's setting:
Ref.~\cite{ding2025simple} requires only $\|A^i\| \le 1$ from a set closed under Hermitian
conjugation, which is strictly more general and is why its Eq.~(1) carries an explicit
$A_S^\dagger \otimes B_E^\dagger$ term. With $A_S$ Hermitian the two terms coincide, which is what makes the
coupling a single weight-2 Pauli and the gate count what it is.

Presets as published, both at Trotter order 2: \texttt{hardware} (the 139-qubit run) is $T = 0.75$,
$\delta t = 0.25$, 3 steps per cycle, $\alpha = 1.75$, $\omega_{\max} = 8$, and a \emph{constant} envelope,
$\sigma = \infty$, which is what Methods~D of Ref.~\cite{farrell2026preparing} specifies for that run:
``With this few Trotter steps, a constant filter function ($\sigma=\infty$) is optimal. Thus, $U(T)$ is
time independent''. Its Methods~A gives the reason, that a time-independent evolution ``is also valid, but
results in worse scaling of the fixed-point error''. \texttt{classical} is $T = 8$, $\delta t = 0.25$
(paper: 0.1, deviation D1 and Sec.~\ref{sec:tuning}), 32 steps, $\alpha = 1$, $\sigma^2 = 1/16$ ($\sigma T = 2$), $\omega_{\max} = 8$. Under
$\sigma \to \infty$ the normalization $\sum_j \delta t f(t_j)^2 = 1$ makes the envelope exactly
$f \equiv 1/\sqrt{T}$, so the constant filter is the pointwise limit of the Gaussian and needs no separate
code path. A third \texttt{weak} preset ($\alpha = 0.5$) exists
and is not used in the grid. The $(n+1)$-qubit channel is simulated exactly as a Kraus map: the
randomization over $(i_S, O_S, \omega)$ is state-independent, so the shot-averaged density matrix evolves
under the averaged channel of Eq.~(2) of Ref.~\cite{ding2025simple}, evaluated exactly over all $3n$ (site, Pauli)
choices and an $n_\omega$-point midpoint quadrature of $g(\omega)$ ($n_\omega = 12$, set by the
\texttt{--n-omega 12} flag since the preset's default is 8: identical to 16, 24, 32 in the fourth decimal at
$\beta = 0.2, 1$, about 8\,\% pessimistic at $\beta = 5$, deviation D3). The Kraus operators are $\beta$-independent, so one channel serves a
whole $\beta$ sweep.

\subsection{The double-bracket thermofield double (arm 2)}

Source: Ref.~\cite{wright2026double}, with the reflection convention and cost recursion from
Ref.~\cite{gluza2026double}. The thermofield double purifies $\rho_\beta$, and imaginary-time evolution for
duration $\beta/2$ carries $|\mathrm{TFD}(0)\rangle$ (the maximally entangled state, whose reduction is the
maximally mixed state) to $|\mathrm{TFD}(\beta)\rangle$. DB-TFD implements that as a double-bracket
flow with the state-dependent generator $[\Psi,H]$, $\Psi = |\psi\rangle\langle\psi|$: \texttt{vanilla}
takes $k$ steps of uniform size $s_i = \beta/(2k)$, and \texttt{poly} applies a degree-$k$ polynomial
$p(H) = \prod(H - z_i)$ one root at a time, each root fixing $(s_i,\theta_i)$ from the measured energy and
variance of the current state (root identity checked to $10^{-12}$).

Compilation: $|\mathrm{TFD}(0)\rangle$ is $n$ Bell pairs, $e^{-itH}$ goes through the shared order-2
Trotter policy, and the reflection $R_0(\theta) = e^{i\theta|0\cdots0\rangle\langle0\cdots0|}$ is a
multi-controlled phase on $L = 2n$ qubits charged at $12L-32$ CNOTs, ancilla-free (64 / 112 / 160 at
$n = 4/6/8$), which is Ref.~\cite{gluza2026double}'s own costing, so the arm is charged at its source's
convention. The V-chain alternative ($12(L-2)+2$ CNOTs with $L-2$ clean ancillas) is, by those two formulas,
never cheaper: the two counts differ by a constant 10 CNOTs at every $L$, so the V-chain costs 74 against
64 CNOTs at $n = 4$, 122 against 112 at $n = 6$ and 170 against 160 at $n = 8$, i.e.\ $15.6$, $8.9$ and
$6.2\,\%$ \emph{more} in two-qubit gates while additionally consuming $L-2$ clean ancillas. The relative
penalty falls with $L$, because the gap is constant while both counts grow linearly, but it does not change
sign at any $L$. Adopting the V-chain here would therefore cost gates and contradict the algorithm's
ancilla-free claim. It is recorded as a side channel rather than used, and any claim that it wins at scale
would need a decomposition whose count is not one of these two. Because
the group commutator has one non-free reflection per step, the cost recursion gives $C_k \sim 3^k$,
reproduced by our own recursion rather than assumed. The state is stored as the $D \times D$ matrix
$M[b_{\text{sys}}, b_{\text{anc}}]$ with $D = 2^n$ (the footprint of one density matrix), so the doubled
register is cheap and this arm's ceiling, $n = 12$, is the exact reference's ceiling rather than its own.

\subsection{The local-driving Lindbladian (arm 3)}

Source: Ref.~\cite{hahn2025towards}. One Lindbladian step: ancillas in $|0\cdots0\rangle$, evolve system
plus bath over $t \in [-T/2, T/2]$ under the interaction-picture Hamiltonian
$\tilde H_{SB}(t) = J\sum_a \{f(t) B_a^\dagger \otimes A_a(t) + f^*(t) B_a \otimes A_a^\dagger(t)\}$ with
$A_a(t) = e^{iHt}A_ae^{-iHt}$, and trace out the ancillas. The filter is the complex shifted Gaussian
$f(t) = \sqrt{2/(\pi\sigma^2)}\exp[-(2/\sigma^2)(t - i\beta/4)^2]$, whose transform is
$\hat f(-\nu) = e^{-\beta\nu/4}e^{-\sigma^2\nu^2/8}$: the entire content of detailed balance is the
imaginary shift $i\beta/4$, which is why the ladder-operator finding of Sec.~\ref{sec:reimpl-ld} carries the consequence it
does. The clock is
$\Delta\tau = J^2$ per cycle. Jump operators $A_i = Y_i$ on every site, $n_B = n_S$.

Because $\tilde H_{SB}$ is an interaction-picture object it cannot be Trotterized naively. Freezing the
drive at pulse midpoints and telescoping the $e^{\pm iHt_j}$ factors gives
$\tilde V \approx e^{+iH(T-\delta t)/2}\,[\prod_j e^{-iV(t_j)\delta t}e^{-iH\delta t}]\,e^{+iH(T-\delta t)/2}$,
i.e.\ $2(T-\delta t) \approx 2T$ of Hamiltonian evolution per cycle, exactly the doubling the paper
attributes to rewinding (the two outer blocks are the rewind). Dropping them leaves the bare lab evolution,
which is the paper's own $K'$.

The drive compiles \emph{exactly}, not approximately: with $A_a$ Hermitian and
$B = |0\rangle\langle 1| = \tfrac12(X+iY)$,
$J\{fB^\dagger\otimes A + f^*B\otimes A^\dagger\} = J\,A\otimes(\mathrm{Re}f\,X_B + \mathrm{Im}f\,Y_B)$, an
axis in the ancilla's $XY$ plane, so one pulse is
$R_Z(\varphi)\,e^{-i\theta A\otimes X_B}\,R_Z(\varphi)^\dagger$ with $\theta = \delta t\,J|f(t_j)|$,
$\varphi = \arg f(t_j)$: three Pauli rotations, two of them weight 1 and free. The arm therefore emits
nothing but Pauli rotations, a measured property, and this is what settles the weight-versus-primitive
question of Sec.~\ref{sec:inversions}.

Step-size policy (rule R2): $T = 6\sigma$, the paper's own value, giving truncation error $10^{-8}$.
The width $\sigma$ is the principal frontier parameter and the main object of the tuning budget. The coupling $J$ sets
$\Delta\tau = J^2$ and the $O(J^4)$ Magnus error and is tuned. The pulse spacing is
$\delta t = \min(\sigma/p,\ \nu_{\text{safety}}\pi/(E_{\max}-E_{\min}))$ with $p = 4$,
$\nu_{\text{safety}} = 0.5$ declared (the second factor is the system--bath arm's Nyquist condition
transplanted, and which factor binds is itself informative). Trotter steps follow the shared order-2 policy
$\text{steps} = \max(1, \lceil|\tau|\|H\|_1/\text{angle\_budget}\rceil)$.

\subsection{Verification}
\label{sec:ourbug}

The same reimplementation process located one error in the harness itself.
\texttt{accounting.\_BASIS\_CHANGE["Y"]} held the inverse of the pair it needed: the compiled block
for a Pauli rotation is $W^\dagger e^{-i\theta Z}W$, so the pre-gate must satisfy $W^\dagger Z W = P$, and
for $Y$ that is $W = \sqrt{X}$, not $\sqrt{X}^\dagger$. Any Pauli string with an odd number of $Y$
factors therefore compiled to $e^{+i\theta P}$. Nothing in the first two arms could expose it: TFIM has no
$Y$, Heisenberg's $YY$ has two sign errors that cancel, DB-TFD's Trotter factors are $ZZ$ and $X$, and the
system--bath arm simulates from the schedule rather than the compiled circuit. The basis change is
single-qubit, hence free in two-qubit accounting, so no gate count reported above changes by a single gate.
The Lindbladian arm's coupling $Y_i \otimes X_a$ has exactly one $Y$, and its meter-equals-schedule test,
which compares the compiled circuit's unitary column by column against the schedule's, failed immediately.
All 63 three-qubit Pauli strings are now checked against \texttt{expm}. A harness that tests only counts
can carry a latent dynamics error indefinitely. The guard is a test that materializes the counted circuit
and simulates it gate by gate.

Beyond the counted-equals-simulated guarantee, the exact KMS-detailed-balanced sampler annihilates
$\rho_\beta$ to $10^{-13}$ relative and satisfies the KMS condition operator by operator at $n = 2,3$ for
$\beta \in \{0.2,1,5\}$, so detailed balance is checked rather than assumed. The closed-form filter
transform is checked against brute-force quadrature. The drive pulse equals
$\mathrm{expm}(-i\,\delta t\,V(t))$ for both ladder conventions. The compiled cycle is simulated gate by
gate by an independent statevector simulator and compared column by column with the unitary the channel
applies. Channels are trace-preserving to $10^{-12}$, and all 63 three-qubit Pauli strings compile to
$e^{-i\theta P}$ as checked against \texttt{expm}.

\section{Deviations from the source papers}
\label{app:deviations}

Each deviation is a field in the corresponding configuration object and is recorded verbatim in every JSON
artifact. Full statements with measured effects are in the stage reports, available on request.

Arm 1 carries seven deviations.
D1 $\delta t = 0.25$ in the \texttt{classical} preset against the source's 0.1: at $n = 4$,
$\beta = 1$ the steady state is 0.0494 at 0.25, 0.0332 at 0.125, 0.0309 at 0.0625, so reported accuracies
are \emph{pessimistic} by about 60\,\% and the gate counts accurate for the $\delta t$ we ran. The
``n/r'' entries of that preset at the two tightest targets are conditional on it (Sec.~\ref{sec:tuning}).
D2 the $(\text{site},\text{Pauli},\omega)$ randomization is averaged exactly rather than sampled,
with no effect on the fixed point.
D3 $g(\omega)$ is a fixed $n_\omega$-point quadrature, a simulator artifact. The grid runs
$n_\omega = 12$, which the controlled sweep shows is converged at $\beta = 0.2$ and $\beta = 1$ (identical
to $n_\omega = 16, 24, 32$ in the fourth decimal) but \emph{not} at $\beta = 5$, where at $n = 4$ it gives
$\td = 0.1134$ against $0.1050$ from $n_\omega \ge 24$. The $\beta = 5$ system--bath accuracies are
therefore pessimistic by about 8\,\%, one-signed and against the system--bath arm, and no gate count moves.
The quadrature is a property of the simulator and not of the protocol.
D4 second-order Trotter, matching the papers.
D5 maximally mixed initial state.
D6 $A_S$ specialized to a Hermitian Pauli, narrower than Ref.~\cite{ding2025simple}'s general
bounded $A^i$ (App.~\ref{app:protocols}). It is what makes the coupling weight 2.
D7 no SWAP routing.
No hyperparameter of this arm is inferred: the \texttt{hardware} envelope is the constant one its
source specifies, and the Gaussian-envelope control runs are kept as \texttt{hardware\_gauss}
rather than used in any table.

Arm 2 carries nine deviations.
D1 two cost models, only one of which is a circuit. This is the headline, $1$--$3.0\times10^6\times$
on TFIM and $165$--$2.1\times10^7\times$ on Heisenberg under second-order product-formula compilation, an
upper bound on the algorithm's own gap (Sec.~\ref{sec:oraclegap}).
D2 group commutator only ($m = 2$), refined by exact sub-stepping rather than by Eq.~(G9) as
printed, so our \texttt{pf} costs are an achievable circuit and not necessarily the best achievable one. A
correct fourth-order formula does exist (Ref.~\cite{chen2022efficient}) and would lower them.
D3 the multi-controlled phase is charged at $12L-32$ CNOTs, ancilla-free, its internal layout not
reproduced (a marker instruction carries the exact logical action, tagged CNOTs carry the charge), so
two-qubit \emph{depth} is not meaningful for this arm and is not reported.
D4 real-time evolution Trotterized. The paper treats $e^{iHt}$ as an oracle.
D5 Chebyshev interpolant on the exact spectral interval (generous, since a real implementation
would need a bound such as $\|H\|_1$: 15 against $\lambda = 9.84$ at $n = 8$).
D6 root ordering a declared, tuned policy. Accuracy is ordering-independent in the oracle model,
cost is not (up to $1.12\times$ at fixed Trotter angle budget).
D7 shot noise not simulated, measurement load carried as a side channel.
D8 the frontier is a minimum over all $(\text{config},\text{iteration})$ points, generous. The
same generosity applies to Arm 3 and with more force, and Sec.~\ref{sec:budgetfrontier} prices both.
D9 Trotter order 2 throughout.

Arm 3 carries ten deviations.
D1 the ladder operator is the one the derivation requires, not the one the display writes
(Sec.~\ref{sec:reimpl-ld}).
D2 $H_B = 0$, the case all the paper's numerics use.
D3 $\tilde V$ discretized into $M$ midpoint pulses. Doubling $\text{points\_per\_sigma}$ from 4 to
8 moves the fixed point by $<3\,\%$ where the Nyquist factor does not already bind and by 0\,\% where it
does.
D4 $e^{\pm iH\tau}$ Trotterized in the \texttt{circuit} model.
D5 $K'$ realized as ``drop the two rewind blocks'' of the interaction-picture telescoping, which
is what the paper describes. It is recorded because the rewind is explicit in our factorization and implicit in
the paper's formula, not because the two disagree (the time-window finding of Sec.~\ref{sec:reimpl-ld}).
D6 jump operators $A_i = Y_i$ (the choice of v2's App.~E, whereas the jump set of the paper's own
numerical case study, its Fig.~3, is not stated),
with $\text{jump\_support} \in \{2,3\}$ swept and strictly worse on both axes.
D7 TFIM at $h = J$ rather than the paper's mixed-field Ising, knowingly adverse to this arm. The
Heisenberg cross-check (Sec.~\ref{sec:heisenberg}) is adverse to it for a second and different reason,
which is why that section reports the mechanism rather than the number.
D8 $\sigma$ tuned per $\beta$ rather than fixed at 0.5, large and favorable, and v2's own
Sec.~IV~D asks for it.
D9 shot noise not simulated. This arm estimates zero Pauli strings, and this side channel is
therefore reported as zero.
D10 Trotter order 2 throughout.

\end{document}